\documentclass[longauth]{aa}

\usepackage[switch]{lineno}
\usepackage{graphicx}
\usepackage{arydshln}
\usepackage{subcaption}
\usepackage{hyperref}
\hypersetup{colorlinks=true,linkcolor=blue,filecolor=blue,urlcolor=blue,citecolor=blue}
\usepackage{appendix}
\usepackage{comment}
\usepackage{anyfontsize}
\usepackage{orcidlink}
\usepackage{multirow}

\usepackage{txfonts}
\makeatletter
\renewcommand*\aa@pageof{, page \thepage{} of \pageref*{LastPage}}
\makeatother

\begin{document}

   \title{Insights from the first flaring activity of a high-synchrotron-peaked blazar with X-ray polarization and VHE gamma rays}
   \titlerunning{Radio to TeV observations of a flare in Mrk~421 during the {IXPE} campaign}
%
\author{\small MAGIC Collaboration:
K.~Abe\inst{1} \and
S.~Abe\inst{2} \orcidlink{0000-0001-7250-3596} \and
J.~Abhir\inst{3}\orcidlink{0000-0001-8215-4377} \and
A.~Abhishek\inst{4} \and
V.~A.~Acciari\inst{5} \orcidlink{0000-0001-8307-2007}\and
A.~Aguasca-Cabot\inst{6} \orcidlink{0000-0001-8816-4920}\and
I.~Agudo\inst{7}  \orcidlink{0000-0002-3777-6182} \and
T.~Aniello\inst{8} \and
S.~Ansoldi\inst{9,43} \orcidlink{0000-0002-5613-7693} \and
L.~A.~Antonelli\inst{8} \orcidlink{0000-0002-5037-9034}
% ...
}
\institute { Japanese MAGIC Group: Department of Physics, Tokai University, Hiratsuka, 259-1292 Kanagawa, Japan
\and Japanese MAGIC Group: Institute for Cosmic Ray Research (ICRR), The University of Tokyo, Kashiwa, 277-8582 Chiba, Japan
\and ETH Z\"urich, CH-8093 Z\"urich, Switzerland
\and Universit\`a di Siena and INFN Pisa, I-53100 Siena, Italy
\and Institut de F\'isica d'Altes Energies (IFAE), The Barcelona Institute of Science and Technology (BIST), E-08193 Bellaterra (Barcelona), Spain
\and Universitat de Barcelona, ICCUB, IEEC-UB, E-08028 Barcelona, Spain
\and Instituto de Astrof\'isica de Andaluc\'ia-CSIC, Glorieta de la Astronom\'ia s/n, 18008, Granada, Spain
\and National Institute for Astrophysics (INAF), I-00136 Rome, Italy
\and Universit\`a di Udine and INFN Trieste, I-33100 Udine, Italy
\and Max-Planck-Institut f\"ur Physik, D-85748 Garching, Germany
\and Universit\`a di Padova and INFN, I-35131 Padova, Italy
\and Croatian MAGIC Group: University of Zagreb, Faculty of Electrical Engineering and Computing (FER), 10000 Zagreb, Croatia
\and Centro Brasileiro de Pesquisas F\'isicas (CBPF), 22290-180 URCA, Rio de Janeiro (RJ), Brazil
\and IPARCOS Institute and EMFTEL Department, Universidad Complutense de Madrid, E-28040 Madrid, Spain
\and Instituto de Astrof\'isica de Canarias and Dpto. de  Astrof\'isica, Universidad de La Laguna, E-38200, La Laguna, Tenerife, Spain
\and University of Lodz, Faculty of Physics and Applied Informatics, Department of Astrophysics, 90-236 Lodz, Poland
\and Centro de Investigaciones Energ\'eticas, Medioambientales y Tecnol\'ogicas, E-28040 Madrid, Spain
\and Departament de F\'isica, and CERES-IEEC, Universitat Aut\`onoma de Barcelona, E-08193 Bellaterra, Spain
\and Universit\`a di Pisa and INFN Pisa, I-56126 Pisa, Italy
\and INFN MAGIC Group: INFN Sezione di Bari and Dipartimento Interateneo di Fisica dell'Universit\`a e del Politecnico di Bari, I-70125 Bari, Italy
\and Armenian MAGIC Group: A. Alikhanyan National Science Laboratory, 0036 Yerevan, Armenia
\and Department for Physics and Technology, University of Bergen, Norway
\and INFN MAGIC Group: INFN Sezione di Torino and Universit\`a degli Studi di Torino, I-10125 Torino, Italy
\and Croatian MAGIC Group: University of Rijeka, Faculty of Physics, 51000 Rijeka, Croatia
\and Universit\"at W\"urzburg, D-97074 W\"urzburg, Germany
\and Technische Universit\"at Dortmund, D-44221 Dortmund, Germany
\and Japanese MAGIC Group: Physics Program, Graduate School of Advanced Science and Engineering, Hiroshima University, 739-8526 Hiroshima, Japan
\and Deutsches Elektronen-Synchrotron (DESY), D-15738 Zeuthen, Germany
\and Armenian MAGIC Group: ICRANet-Armenia, 0019 Yerevan, Armenia
\and Croatian MAGIC Group: University of Split, Faculty of Electrical Engineering, Mechanical Engineering and Naval Architecture (FESB), 21000 Split, Croatia
\and Croatian MAGIC Group: Josip Juraj Strossmayer University of Osijek, Department of Physics, 31000 Osijek, Croatia
\and Finnish MAGIC Group: Finnish Centre for Astronomy with ESO, Department of Physics and Astronomy, University of Turku, FI-20014 Turku, Finland
\vfill\null
\and University of Geneva, Chemin d'Ecogia 16, CH-1290 Versoix, Switzerland
\and Saha Institute of Nuclear Physics, A CI of Homi Bhabha National Institute, Kolkata 700064, West Bengal, India
\and Inst. for Nucl. Research and Nucl. Energy, Bulgarian Academy of Sciences, BG-1784 Sofia, Bulgaria
\and Japanese MAGIC Group: Department of Physics, Yamagata University, Yamagata 990-8560, Japan
\and Finnish MAGIC Group: Space Physics and Astronomy Research Unit, University of Oulu, FI-90014 Oulu, Finland
\and Japanese MAGIC Group: Chiba University, ICEHAP, 263-8522 Chiba, Japan
\and Japanese MAGIC Group: Institute for Space-Earth Environmental Research and Kobayashi-Maskawa Institute for the Origin of Particles and the Universe, Nagoya University, 464-6801 Nagoya, Japan
\and Japanese MAGIC Group: Department of Physics, Kyoto University, 606-8502 Kyoto, Japan
\and INFN MAGIC Group: INFN Roma Tor Vergata, I-00133 Roma, Italy
\and Japanese MAGIC Group: Department of Physics, Konan University, Kobe, Hyogo 658-8501, Japan
\and also at International Center for Relativistic Astrophysics (ICRA), Rome, Italy
% ...
}

   \date{Received XX XX, 2024; accepted XX XX, 2024}

 
  \abstract
   {Blazars exhibit strong variability across the entire electromagnetic spectrum, including periods of high-flux states commonly dubbed as flares. The physical mechanisms in blazar jets responsible for flares remain to date poorly understood.}
   { We aim to better understand the emission mechanisms during blazar flares using X-ray polarimetry and broadband observations from the archetypical TeV blazar Mrk~421, which can be studied with higher accuracy than other blazars that are dimmer and/or located farther away.}
   { We study a flaring activity from December 2023 that was characterized from radio to very-high-energy (VHE; \mbox{E $>0.1$\,TeV)} gamma rays with MAGIC, \textit{Fermi}-LAT, \textit{Swift}, \textit{XMM-Newton} and several optical and radio telescopes. These observations included, for the first time for a gamma-ray flare of a blazar, simultaneous X-ray polarization measurements with IXPE, besides optical and radio polarimetry data. We quantify the variability and correlations among the multi-band flux and polarization measurements, and describe the varying broadband emission within a theoretical scenario constrained by the polarization data.}
  {We find substantial variability in both X-rays and VHE gamma rays throughout the campaign, with the highest VHE flux above 0.2\,TeV occurring during the IXPE observing window, and exceeding twice the flux of the Crab Nebula. However, the VHE and X-ray spectra are on average softer, and the correlation between these two bands weaker that those reported in previous flares of Mrk~421. IXPE reveals an X-ray polarization degree significantly higher than that at radio and optical frequencies, similar to previous results on Mrk~421 and other high-synchrotron-peaked blazars. Differently to past observations, the X-ray polarization angle varies by $\sim$100$^\circ$ on timescales of days, and the polarization degree changes by more than a factor 4. The highest X-ray polarization degree, analyzed in 12\,hrs time intervals, reaches $26\pm2\%$, around which an X-ray counter-clockwise hysteresis loop is measured with \textit{XMM-Newton}. It suggests that the X-ray emission comes from particles close to the high-energy cutoff, hence possibly probing an extreme case of the Turbulent Extreme Multi-Zone model for which the chromatic trend in the polarization may be more pronounced than theoretically predicted. We model the broadband emission with a simplified stratified jet model throughout the flare. The polarization measurements imply an electron distribution in the X-ray emitting region with a very high minimum Lorentz factor ($\gamma'_{\rm min}\gtrsim10^4$), which is expected in electron-ion plasma, as well as a variation of the emitting region size up to a factor of three during the flaring activity. We find no correlation between the fluxes and the evolution of the model parameters, which indicates a stochastic nature of the underlying physical mechanism that likely explains the lack of a tight X-ray/VHE correlation during this flaring activity. Such behaviour would be expected in a highly turbulent electron-ion plasma crossing a shock front.}
   {}
   \keywords{BL Lacertae objects:  individual (Markarian 421)   galaxies:  active   gamma rays:  general radiation mechanisms:  nonthermal  X-rays:  galaxies}

   \maketitle
%


\section{Introduction} \label{sec:intro}

Blazars are a subclass of active galactic nuclei (AGNs) characterized by a powerful relativistic plasma jet \citep{2019ARA&A..57..467B}. The jet's axis is aligned at a small angle with the observer’s line of sight, leading to strong relativistic aberration of the observed radiation. Their broadband emission is dominated by non-thermal radiation from the jet that goes from radio to very-high-energy (VHE; $>0.1$\,TeV) gamma rays. \par 

One of the key features of their emission is a high degree of variability observed over the full spectrum from years to hour timescales \citep[see e.g.][]{2008ApJ...677..906F, 2008ApJ...689...79C}. During high-flux states, commonly dubbed as flares, extreme flux variations up to an order of magnitude on the timescale of minutes have also been reported \citep{2007ApJ...669..862A,2007ApJ...664L..71A}. Flares are not only accompanied by strong flux changes but also by large spectral variations \citep{1998ApJ...492L..17P}, hence implying that particle acceleration mechanisms play a central role in the origin of the variability. However, all those phenomena, and in particular the underlying processes which accelerate particles to highly-relativistic energies, are still a topic of debate.\par 

Among the acceleration mechanisms commonly considered for blazars there are shock acceleration \citep[][]{1978ApJ...219..392M, 2019MNRAS.485.5105C} and magnetic reconnection \citep{2014ApJ...783L..21S}. For the first, the source of acceleration is the interaction with a collisionless shock wave, where turbulences in the magnetic field of the plasma are responsible for the multiple crossings at the shock front, leading to an efficient energy gain. In the case of magnetic reconnection, it is the magnetic field lines themselves that are the root of the acceleration. Through instabilities in the magnetic field, field lines of opposite polarity can reconnect with each other and transfer magnetic energy to kinetic energy efficiently. Which of these two mechanisms is dominating in blazars is still unknown. It is also dependent on where we expect the observed emission taking place in the jet due to different dependencies on the magnetization of the jet.\par 

An important tool to probe acceleration mechanisms are polarization measurements since they directly relate to the structures of the magnetic fields at the emission sites, which in return lead to constrains on the possible acceleration mechanisms \citep{2021Galax...9...37T}. As discussed in \cite{2020MNRAS.498..599T, 2022A&A...662A..83D}, in the scenario of shock acceleration, the polarization degree is expected to show a strong chromatic behaviour. The polarization increases with energy because the highest-energy particles are located closer to the shock front where they are freshly accelerated, and where the magnetic field is more ordered as it gets compressed. The polarization degree is expected to be slowly variable, and the angle parallel to the shock normal (i.e. parallel to the jet axis in case of a shock normal aligned with the jet). In case of a highly turbulent plasma crossing a shock front, both the polarization degree and angle can exhibit strong variability dictated by the stochastic nature of the magnetic field in the plasma cells crossing the shock. The average polarization angle is nevertheless expected to be roughly aligned with the shock normal due to the partial ordering and compression of the field by the shock. In this context, \citet{2014ApJ...780...87M,2017Galax...5...63M} also developed the ``Turbulent Extreme Multi-Zone Model for Blazar Variability'' (TEMZ), in which a turbulent plasma crosses a standing shock (e.g. recollimation shock). This model predicts a chromatic behaviour of the polarization driven by the turbulent nature of the plasma, and is consistent with radio-to-optical polarization behaviors seen in several blazars \citep[e.g.][]{2018A&A...619A..45M}. For magnetic reconnection, the polarization degree and angle can show fast (below the light-crossing time of the reconnection layer) and chaotic variability driven by the complex geometry of the current sheets \citep{2022ApJ...924...90Z}.\par 

Until recently, polarization observations of blazar jets were only available in the optical and radio wavebands. However, since the end of 2021, the Imaging X-ray Polarimetry Explorer \citep[IXPE;][]{2022HEAD...1930101W} is providing us with linear polarization observations in the 2-8\,keV band. For high synchrotron peaked blazars (HSPs), which have a synchrotron spectral energy distribution (SED) peaking above $10^{15}$\,Hz \citep{2010ApJ...716...30A}, the IXPE energy regime is measuring the synchrotron emission from the most energetic, freshly accelerated electrons. Therefore, X-ray polarization measurements directly probe the conditions close to the acceleration regions for HSPs, while optical and radio observation are most probably connected to regions further downstream the jet due to the longer cooling timescale \citep{2021Galax...9...37T}. Combining radio-to-X-ray polarization observations provides a unique opportunity to disentangle the acceleration mechanisms in blazars and thus better understand the origin of flaring events.\par 

During its first years of operations, the HSPs observed by IXPE had been probed in non-flaring activity. In 2022, IXPE announced the first detection of X-ray polarization from a blazar \citep{2022Natur.611..677L}, the archetypal HSP Markarian~501 (hereafter Mrk~501), revealing significantly higher polarization degrees than in the optical or radio regimes, but similar polarization angles. The same was found for another archetypal HSP, Markarian~421 \citep[hereafter Mrk~421; ][]{2022ApJ...938L...7D,2023NatAs...7.1245D}. This indicated that (at least) in non-flaring activity shock acceleration in an energy stratified jet is preferred. This scenario is further supported by the results derived with the data from the multi-instrument campaigns organized for these two sources during the year 2022 \citep{2024A&A...684A.127A,2024A&A...685A.117M}. \par 

In December 2023, another IXPE observation of Mrk~421 was performed lasting from December 6 (MJD~60284) to December 21  (MJD~60300), around which we organized a dense monitoring campaign from radio to VHE. This time, the source was found in a flaring state reaching VHE fluxes of more than twice the one of the Crab Nebula, as well as in a bright X-ray state compatible with prominent archival flares \citep{2015A&A...578A..22A}. In this work, we present the first broadband study from radio to VHE of a gamma-ray flare of a blazar in combination with simultaneous radio-to-X-ray polarization measurements. For this study, we have coordinated and combined extensive observation by the Florian Goebel Major Atmospheric Gamma Imaging Cherenkov (MAGIC) telescopes \citep{aleksic:2016}, the Large Area Telescope (LAT) on board \textit{Fermi Gamma-ray Space Telescope} \citep[\textit{Fermi}-LAT;][]{2009ApJ...697.1071A,2012ApJS..203....4A},  the \textit{Neil Gehrels Swift} Observatory \citep[\textit{Swift};][]{2004ApJ...611.1005G}, the X-ray Multi-Mirror Mission (\textit{XMM-Newton}) and various ground-based telescopes to cover the optical and radio frequencies. 

The paper is structured as follows. The observations and data processing are described in Sect.~\ref{sec:data_analysis}. The multi-wavelength behaviour and polarization characteristics during the campaign are discussed in Sect.~\ref{sec:characterization_MWL}, and in Sect.~\ref{sec:MWL_correlation} we characterize the intra-band correlations. In Sect.~\ref{sec:modelling}, we perform the modelling of the broadband emission throughout the IXPE window using constraints from the observed multi-wavelength polarization, and the discussion and summary are presented in Sect.~\ref{sec:discussion}.

\section{Observations and data processing} \label{sec:data_analysis}

Most of the observations were carried out by the same instruments and using the same data processing as for the multi-wavelength campaigns recently published in \citet{2024A&A...684A.127A} and \citet{2024A&A...685A.117M}. We refer the reader to the latter works for more details, but we provide below a summary for completeness.\par 

Observations in the VHE band were performed by the MAGIC telescopes at the Observatorio del Roque de los Muchachos (ORM) on the Canary Island of La Palma, Spain. The data were analyzed with the standard procedures and the MAGIC Analysis and Reconstruction Software \citep[MARS;][]{zanin2013, aleksic:2016} software package as in \citet{2024A&A...684A.127A}. We extracted daily light curves for the full campaign in the 0.2-1\,TeV and >1\,TeV bands. Moreover, a deep MAGIC exposure of 4.3\,hours (15.6\,ks) was organized during the multi-hour long observations from \textit{XMM-Newton} that took place on December 13 2023 (MJD~60291), which was used to derive an intranight light curve above 0.4\,TeV in 25-minutes intervals. The latter minimum energy is higher than the threshold adopted for the daily light curve because, on that night, the MAGIC observations were performed at zenith angles from $\approx10^\circ$ to $\approx60^\circ$. At zenith angles close to $\approx60^\circ$, the threshold of MAGIC is $\approx350$\,GeV. Hence, using an increased threshold of 0.4\,TeV allows us to include all the 25-min bins, even those with data taken at the highest zenith angles up to $60^\circ$. Nightly SEDs and spectral parameters were obtained with a forward folding method. Since for the long exposure night, a log-parabola model was preferred by more than 3$\sigma$, it is employed for all nights to ease comparability.\par

For the high-energy (HE) gamma rays, data from \textit{Fermi}-LAT were obtained and analyzed following the same procedure as in \citet{2024A&A...685A.117M}. We produced a light curve with a binning of 3 days and SEDs centered around each MAGIC observation also integrated over 3 days. A simple power-law model is used to produce the SEDs. We note that, when using the 15 days considered in this study (December 6-21), a log parabola is not preferred (with respect to a power law) by more than 3 $\sigma$.\par  

A dense X-ray monitoring campaign with the \textit{Swift}-XRT telescope \citep{2005SSRv..120..165B} was organized to support the MAGIC observations. A special effort was put to schedule the observations simultaneously to the MAGIC observations. The data were reduced as in \citet{2024A&A...684A.127A} using an updated version of the XRTDAS software package (v.3.7.0) developed by the ASI Space Science Data Center\footnote{\url{https://www.ssdc.asi.it/}} (SSDC), released by the NASA High Energy Astrophysics Archive Research Center (HEASARC) in the HEASoft package (v.6.32.1). We extracted fluxes in the 0.3-2\,keV and 2-10\,keV bands by fitting a log-parabola model assuming an hydrogen column density fixed to $N_{\rm H}=1.34\times 10^{20}$\,cm$^{-2}$ \citep[][]{2016A&A...594A.116H}. The same $N_{\rm H}$ will be used throughout this work. In the vast majority of the cases (>97\% of the fits), the log-parabola model is preferred over the power-law model with a significance above $3\sigma$.\par

On December 13 2023 (MJD~60291) we performed a multi-hour long observation with the \textit{XMM-Newton}, which carries on board several coaligned X-ray instruments. One of them is the European Photon Imaging Camera (EPIC), consisting of the Metal Oxide Semiconductor cameras \citep[EPIC-MOS;][]{2001A&A...365L..27T} and the pn junction camera \citep[EPIC-pn;][]{2001A&A...365L..18S} that both operate in the 0.2-10\,keV band. In this work, we focus on the data from the EPIC-pn camera considering its better sensitivity that allows us to resolve spectral changes on short timescale, which is of importance for this work (see Sect.~\ref{sec:spectral_evolution}). Within instrumental systematics, both the EPIC-pn and EPIC-MOS show consistent results. The data were taken in TIMING mode with the THICK filter, for an overall exposure of 16.9\,ks (4.6\,hrs) after good time interval screening. We extracted the source and background spectra following the same analysis procedures as in \citet{2021A&A...655A..48D, 2024A&A...684A.127A}. Due to a count rate close to the threshold above which pile-up may occur, we removed the central column around the source to suppress any potential pile-up artifacts. The fluxes in the 0.3-2\,keV and 2-10\,keV bands were computed by fitting a log-parabola model, which is significantly preferred over a simple power law.\par

The IXPE telescope \citep{2022HEAD...1930101W} is the first instrument capable of resolving the X-ray polarization degree and angle in blazars. In December 2023, four observations took place, from December 6 (MJD~60284) to December 21 (MJD~60300) for a total exposure of 514\,ks spread over four observations. The data processing was performed using the \texttt{ixpeobssim} software, version 30.6.3 \citep{2022SoftX..1901194B}. As in \citet{2023_IXPE_Mrk421}, the $I$, $Q$ and $U$ spectra were determined in the 2-8\,keV band and the \texttt{pcube} algorithm was used to obtain polarization angle and degree. We refer the reader to \citet{2023_IXPE_Mrk421} for more details on the source and background region selection. The data were binned into $\approx12$\,hrs intervals, for which a significant detection of the X-ray polarization is obtained in all intervals, revealing a strong polarization variability. To investigate variability on shorter timescale, the data were also binned into 6\,hrs intervals, and further down to 3\,hrs intervals contemporaneous to the long \textit{XMM-Newton} observations.\par

In the UV band, we analyzed the \textit{Swift}-UVOT images in the W1, M2 and W2 filters, for the observations in the time interval of interest. We applied a reduction and data analysis procedure similar to those of \citet{2024A&A...684A.127A}. The identical HEAsoft software version was used for the aperture photometry task, as well as the same CALDB release to apply the standard calibrations \citep{2011AIPC.1358..373B}. We included calibrations systematic errors to the magnitude statistical uncertainties \citep{2008MNRAS.383..627P, UVOT_CALDB1, UVOT_CALDB2}. However, various observations in this time interval are affected by attitude instabilities, due to increased noise to one of the spacecraft gyroscopes. We checked carefully image photometry, executing photometry in few cases to single "exposure slices" to recover some observations, and we ended discarding about 1/6th of the observations.\par

Optical photometry and polarimetry observations in the R-band were performed by the 2.2\,m telescope of the Calar Alto Observatory as part of the Monitoring AGN with Polarimetry at the Calar Alto Telescopes \citep[MAPCAT\footnote{\url{https://home.iaa.csic.es/~iagudo/_iagudo/MAPCAT.html}};][]{mapcat_program}, the 1.5\,m (T150) and the 0.9\,m (T090) telescopes at the Sierra Nevada Observatory. We also make use of observations from the Nordic Optical Telescope at the ORM \citep[NOT;][]{2018A&A...620A.185N}, the KANATA telescope \citep[Higashi-Hiroshima observatory, Japan][]{1999PASP..111..898K, 2014SPIE.9147E..4OA}, the Liverpool Telescope \citep{2016SPIE.9908E..4IJ, 2020MNRAS.494.4676S} and the Boston University’s Perkins telescope (Perkins Telescope observatory, Flagstaff, AZ). Finally, we obtained R-band photometry data from the Tuorla blazar monitoring program using the 80\,cm Joan Oró Telescope (TJO) at Montsec Observatory, Spain. The details of the data analysis an reduction for the different telescopes can be found in \citet{2024arXiv240601693K, 2024A&A...684A.127A, 2024A&A...685A.117M, 2022Natur.611..677L, 2018A&A...620A.185N, mapcat_osn_data, mapcat_pipeline}. Additional R-band data were acquired with a Moravian G4-16000 CCD camera equipped with Bessel filters from Chroma Technology at the 0.5m PlaneWave CDK astrograph of the Hans-Haffner-Sternwarte in Hettstadt, Germany, as part of the long-term AGN observation program of the science laboratory for students at the Friedrich-Koenig-Gymnasium (FKG), the University of W\"urzburg, and the TU Dortmund University. To extract the fluxes, photometric settings and comparison stars were taken from the Glast AGILE Support Program (GASP) list of the Whole Earth Blazar Telescope\footnote{\url{https://www.oato.inaf.it/blazars/webt/}}. The polarization data were corrected for the contribution of the host galaxy with the same method as \citet{2016A&A...596A..78H} and using the host fluxes reported in \citet{2007A&A...475..199N}. In order to build broadband SEDs, the flux densities were also corrected for a galactic extinction of 0.033 mag according to the NASA/IPAC Extragalactic Database (NED)\footnote{\url{https://ned.ipac.caltech.edu/}}.\par

In the radio band, the flux density and polarization were measured in the mm band (225.5\,GHz) by the SMA \citep{Ho2004} telescope within the framework of the \textit{SMA Monitoring of AGNs with POLarization} (SMAPOL) program \citep{Myserlis_prep} and the data reduction was performed as in \citet{2024A&A...685A.117M}. We also obtained observations in the cm range (4.85\,GHz, 10.45\,GHz and 14.25\,GHz) thanks to the \textit{Monitoring the Stokes Q, U, I and V Emission of AGN jets in Radio} (QUIVER) program \citep{2018A&A...609A..68M, 2003A&A...401..161K} using the Effelsberg 100 m telescope. We refer the reader to \citet{2018A&A...609A..68M} for more details on the analysis methods.

\begin{figure*}
\centering
\includegraphics[width=0.95\textwidth]{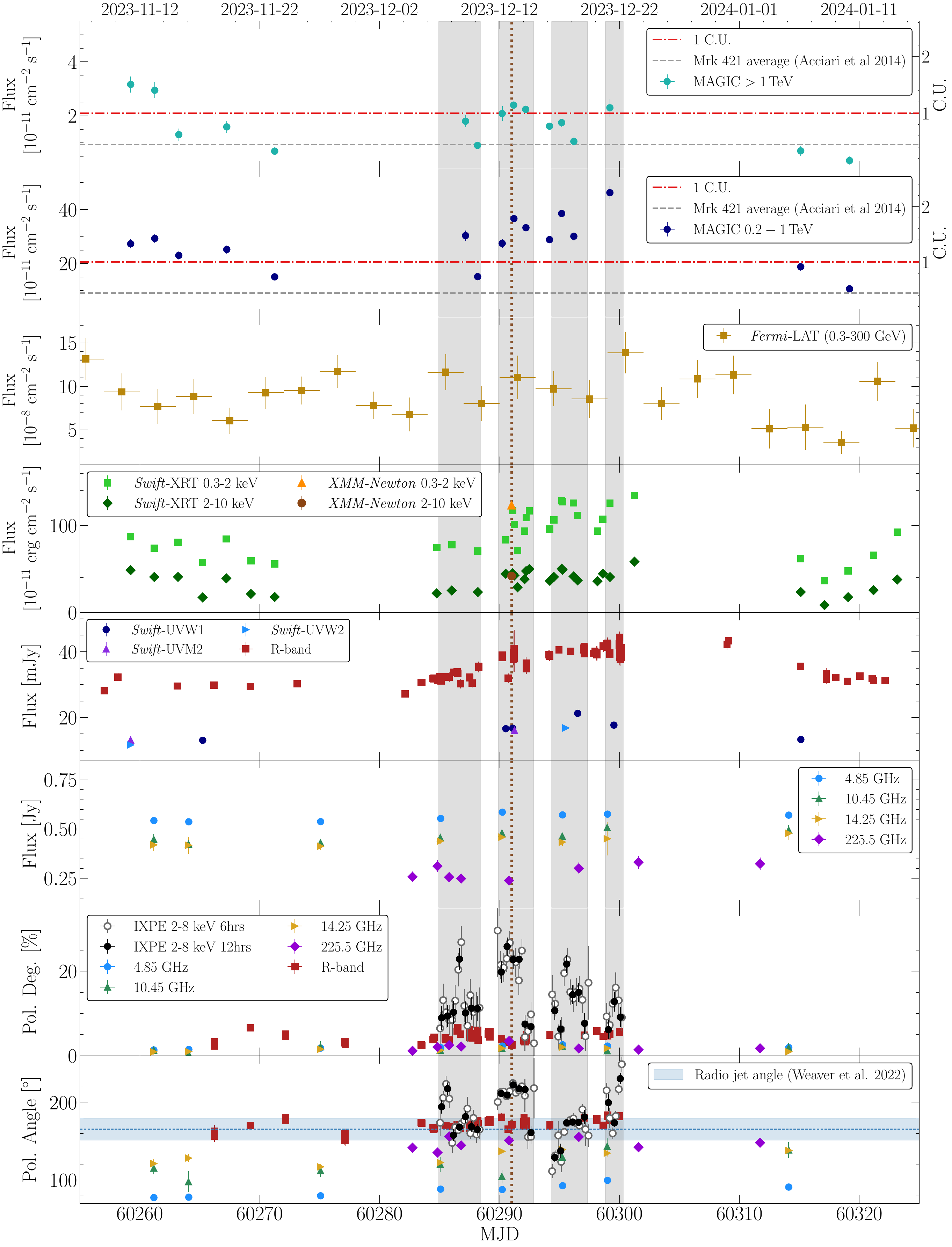}
\caption{\scriptsize Multi-wavelength light curve between November 7 2023 (MJD~60255) and January 16 2024 (MJD~60325). The red line marks the \textit{XMM-Newton} observation taking place on December 13 2023 (MJD~60291), and the vertical grey bands highlight the IXPE observing windows. Top to bottom: MAGIC fluxes in the $>$1\,TeV (first panel) and 0.2-1\,TeV (second panel) bands with nightly bins (the typical nightly exposure is $\approx40$\,min). The grey-dashed line depicts the average flux of Mrk~421 from \citet{2014APh....54....1A}; \textit{Fermi}-LAT fluxes in 3-day bins; X-ray fluxes binned per observation including \textit{Swift}-XRT and \textit{XMM-Newton}; \textit{Swift}-UVOT and optical R-band data using the telescopes listed in Sect.~\ref{sec:data_analysis}; Radio data from the SMAPOL and QUIVER programs; polarization degree \& polarization angle observations in the X-rays from IXPE (in black and empty grey markers the data are binned over 12\,hrs and 6\,hrs bins, respectively), the optical R-band and the radio band. In the polarization angle panel, we plot with a horizontal blue band the average direction angle determined at 43\,GHz by \citet{2022ApJS..260...12W}. See text in Section~\ref{sec:characterization_MWL} for further details.}
\label{fig:longterm_LC}
\end{figure*}

\section{Characterization of the broadband emission and polarization behavior} \label{sec:characterization_MWL}
Fig.~\ref{fig:longterm_LC} depicts the multi-wavelength light curves from the radio up to the VHE gamma-ray range, that span from November 7 2023 (MJD~60255) and January 16 2024 (MJD~60325). These observations were performed around the long IXPE observation taking place from December 6 2023 (MJD~60284) to December 22  2023 (MJD~60300).

\begin{figure}
  \centering
  \includegraphics[width=\columnwidth]{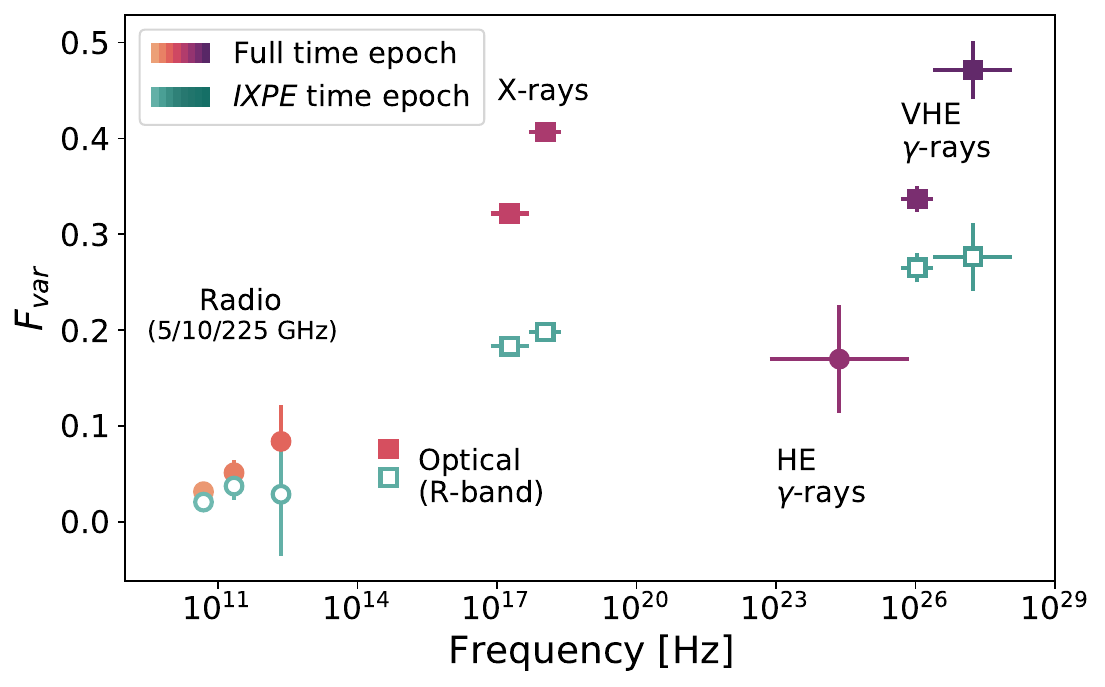}
 \caption{Fractional variability, $F_{var}$, for the light curves displayed in Fig.~\ref{fig:longterm_LC}. Nightly bins are used for MAGIC, 3-day bins for \textit{Fermi}-LAT, and for all other wavebands the single observations without further binning. The filled orange to violet markers depict $F_{var}$ when data over the full campaign are considered and the open turquoise markers only includes measurements during the IXPE time window. For \textit{Fermi}-LAT, $F_{var}$ can not be computed during the IXPE window due to a variability level below the statistical uncertainties, leading to a negative value in the square root of the numerator of $F_{var}$ \citep[see ][]{2003MNRAS.345.1271V}.}
 \label{fig:fvar}
\end{figure}

\begin{figure}
  \centering
  \includegraphics[width=1\columnwidth]{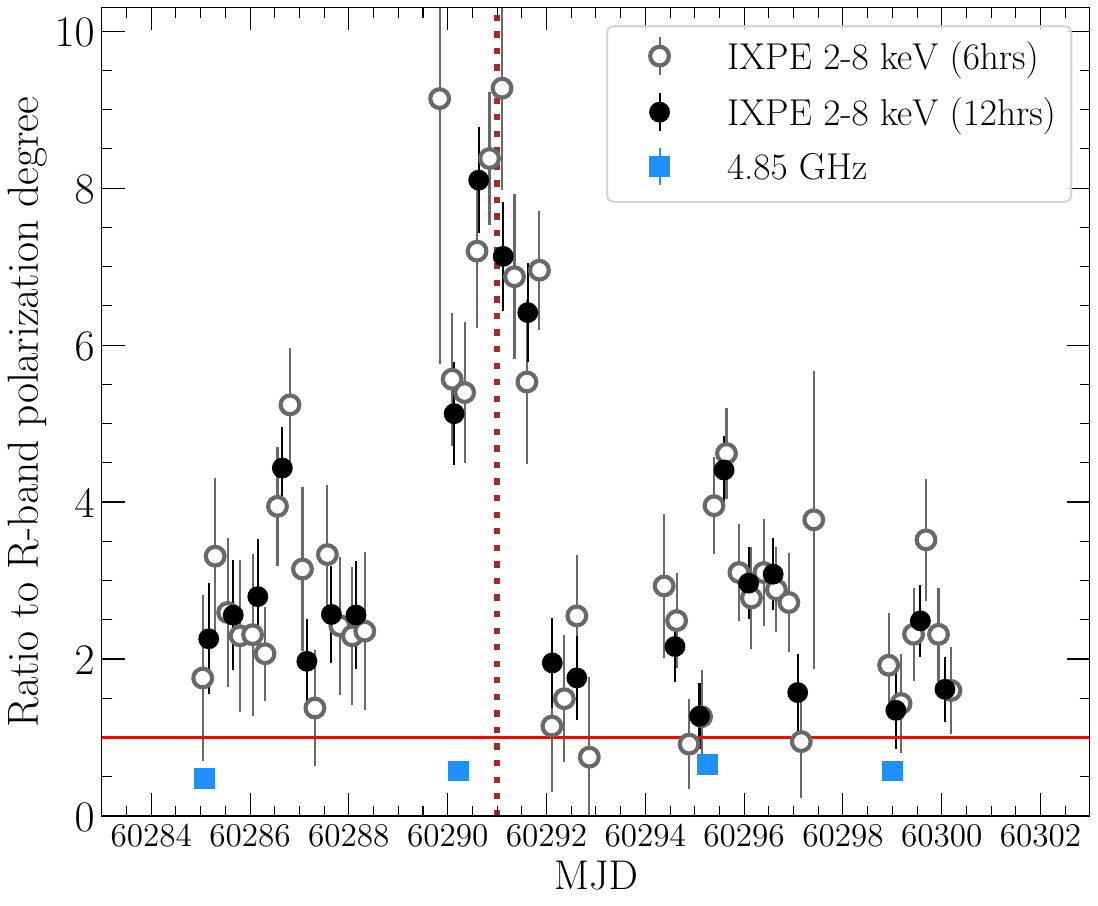}
 \caption{Ratio of the X-ray (2-8\,keV) and radio (4.85\,GHz) polarization degree to the one measured in the R-band during the IXPE observing window ($P_{\rm X-ray, deg}/P_{\rm R-band, deg}$ and $P_{\rm radio, deg}/P_{\rm R-band, deg}$, respectively). The vertical dashed line marks the date of the simultaneous \textit{XMM-Newton}/MAGIC long exposure.}
 \label{fig:pdeg_ratios}
\end{figure}

The MAGIC light curves in the top two panels show that Mrk~421 was in a high VHE state during most of the observation epoch. Notably, two flaring periods exceeding 1 Crab Nebula units\footnote{The flux of the Crab Nebula used in this work is obtained by integrating the spectrum from the Crab Nebula in \citet{aleksic:2016}} (C.U.; see red dashed line in Fig.~\ref{fig:longterm_LC}) can be seen, one in November 2023 and another one in December 2023. For the flare in December 2023, which is simultaneous to the IXPE observations, the 0.2-1\,TeV flux reached more than 2\,C.U. towards the end of the flare. We note that the average flux of Mrk~421 at those energies, plotted as a grey dashed line in Fig.~\ref{fig:longterm_LC}, is around 0.45\,C.U. \citep[][]{abdo:2011, 2014APh....54....1A}, which is more than a factor 4 smaller. The highest VHE flux levels in December 2023 are 4 to 8 times brighter than during the first IXPE observations of Mrk~421 in 2022 where the simultaneous 0.2-1\,TeV flux ranged from 0.25 to 0.5\,C.U. \citep{2024A&A...684A.127A}.

The \textit{Swift}-XRT light curves reveal a bright X-ray state, in particular during the IXPE window. The 0.3-2\,keV fluxes are regularly above $10^{-9}$\,erg cm$^{-2}$ s$^{-1}$, which is higher by at least a factor 2 when considering previous campaigns with close-to-average X-ray states \citep[see e.g.][]{2021A&A...655A..89M}. The 0.3-2\,keV fluxes are in fact comparable to the bright flare of March 2010 \citep{2015A&A...578A..22A}. Differently from the 0.3-2\,keV band, the 2-10\,keV fluxes are closer to average values ($\approx 0.4\times 10^{-9}$\,erg cm$^{-2}$ s$^{-1}$), if one refers to the 6-month campaign of 2017 discussed in \citet{2021A&A...655A..89M}. This difference between the 0.3-2\,keV and 2-10\,keV fluxes relative to the average states from archival data points towards a softer X-ray spectrum than usually observed. This is confirmed in the spectral study presented in Sect.~\ref{sec:spectral_evolution}. In the other wavebands, radio,  optical, UV and HE gamma rays, the fluxes show an absence of strong variability. The emission is comparable to the typical state of Mrk~421 \citep{abdo:2011}, as well as the one noticed in 2022 during the first IXPE observations of Mrk~421.\par

The fractional variability ($F_\text{var}$) is reported in Fig.~\ref{fig:fvar} for all the available wavebands with sufficient number of observations to be representative of the campaign. For instance, the UV bands are excluded because of the very low number of measurements, as well as the highly non-uniform distribution of the observations, for any single UV band from  \textit{Swift}-UVOT. $F_\text{var}$ quantifies the variability strength following Eq.~10 in \citet{2003MNRAS.345.1271V}, and the corresponding (statistical) uncertainty is estimated following \citet{2008MNRAS.389.1427P}. For the full time period, the strongest variability is seen in the VHE gamma rays and the X-rays, with an increased variability ($F_\text{var}>0.5$) in the higher sub-energy range in these two bands ($>1$\,TeV and 2-10\,keV). During the IXPE window (shown with green markers), the VHE gamma rays and X-ray $F_\text{var}$ is $\approx$ 0.2-0.3, and it is very similar between the two sub-energy bands in the X-ray and VHE regimes. This is different from the strong chromatic trend observed over the full campaign. The lower variability and essentially achromatic behaviour of the fractional variability during the IXPE observations may be indicative of a different state. However, we cannot exclude that this difference is due to the shorter time interval that is being considered (15 days vs 70 days), hence reducing the probability to catch flux variability in the source.

\begin{figure*}[h!]
    \centering
    \begin{subfigure}[b]{1\textwidth}
        \centering
        \includegraphics[width=\textwidth]{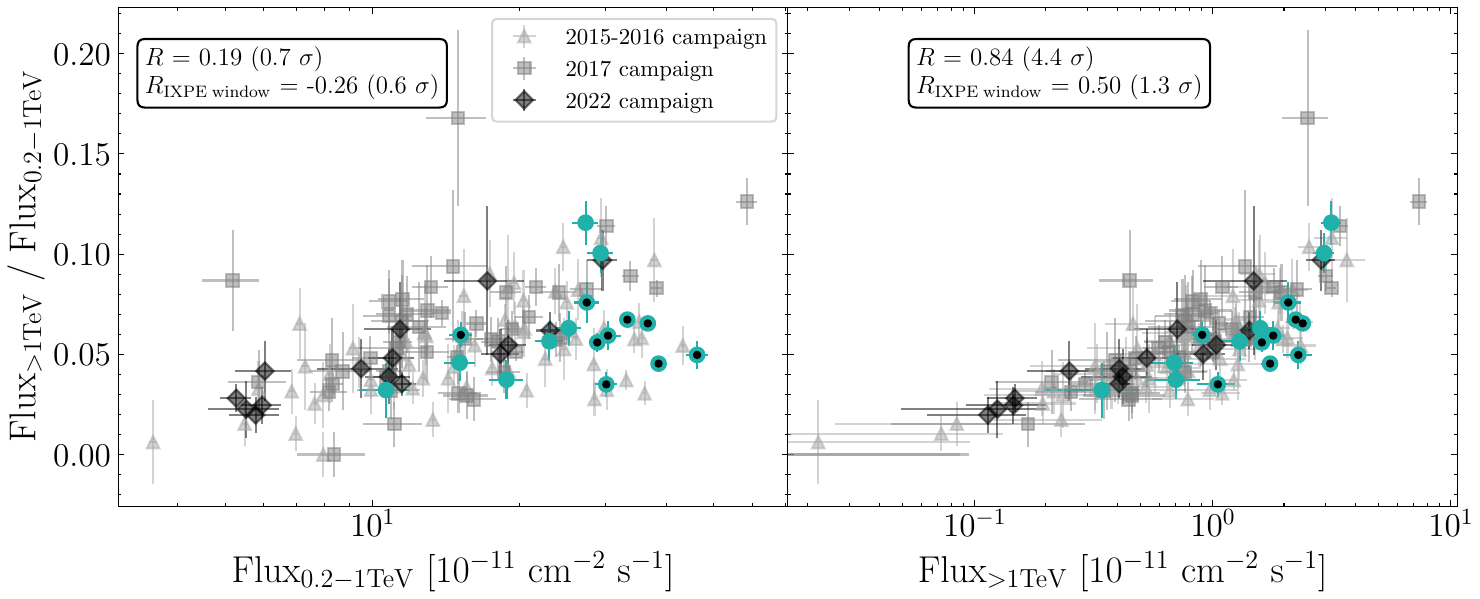}
    \end{subfigure}
    \begin{subfigure}[b]{1\textwidth}
        \centering
        \includegraphics[width=0.98\textwidth]{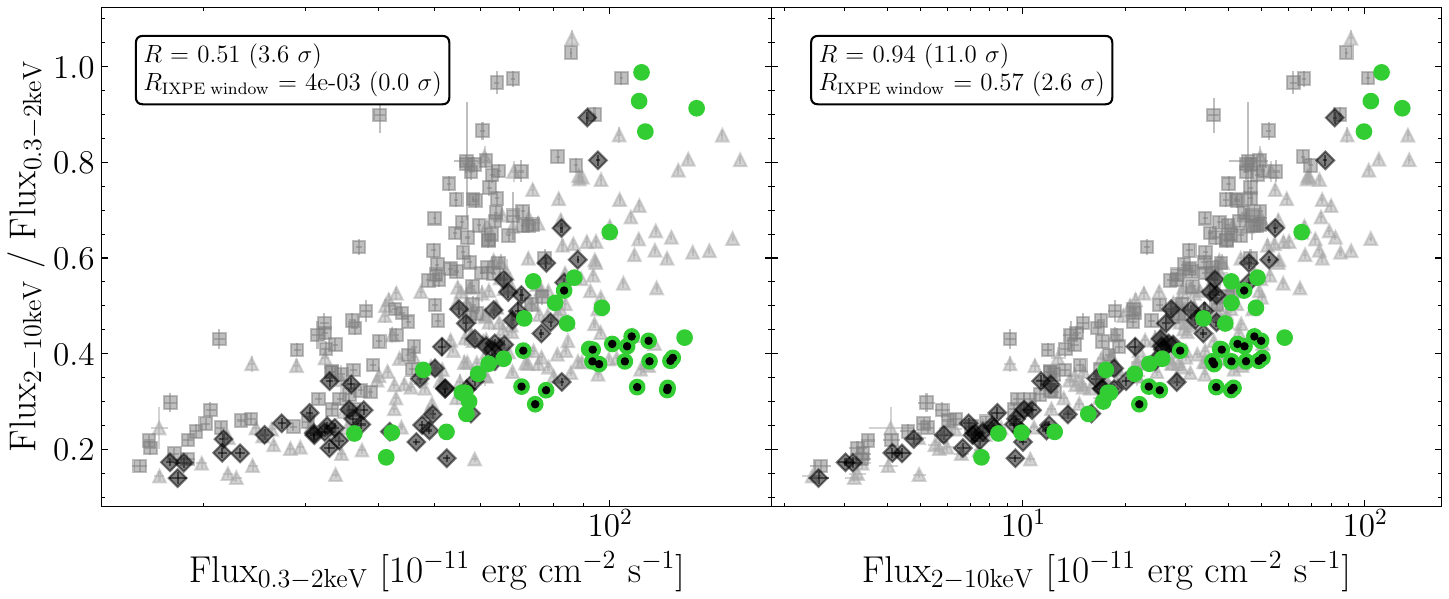}
    \end{subfigure}

    \caption{Hardness ratio in the VHE (upper panels) and X-rays (lower panels) bands. The VHE hardness ratio is defined as the ratio of the $>1$\,TeV to the 0.2-1\,TeV fluxes, while in the X-ray it is the ratio of the 2-10\,keV to the 0.3-2\,keV fluxes. They are plotted versus the 0.2-1\,TeV and $>1$\,TeV bands, while in the X-ray they are plotted versus the 0.3-2\,keV and 2-10\,keV bands. Colorful circular markers correspond to measurements from the campaign under study, and highlights data simultaneous to the IXPE window with black-filled markers. Archival measurements from the extensive multi-instrument observing campaigns in the years 2015-2016 \citep{2021MNRAS.504.1427A}, 2017 \citep{2021A&A...655A..89M}, and 2022 \citep{2024A&A...684A.127A} are plotted with grey triangular, grey squared and diamond black markers, respectively. } 
    \label{fig:HR_xray_vhe}
\end{figure*}

\subsection{Multi-wavelength polarization evolution} 
\label{sec:pol_behavior}
In addition to the multi-wavelength fluxes, polarization measurements in the radio, optical and X-ray bands are shown in the two bottom panels of Fig.~\ref{fig:longterm_LC}. In the different radio bands, the polarization degree varies by about $\pm0.5 \times \overline{P}_{\rm radio, deg.}$, with a mean value of $\overline{P}_{\rm radio, deg.} \approx 2\% $. A similar degree of variability is found in the R-band but the mean value is higher, $\overline{P}_{\rm R-band, deg.}\approx4.5$\%, both for the full campaign as well as during the IXPE observation window. The variability in the X-ray polarization degree is much stronger. The IXPE polarization degree binned over 12\,hrs intervals (black markers) show variations by more than a factor 4 between the minimum ($6\pm2$\%) and maximum ($26\pm2$\%). Over the 6\,hrs intervals (grey empty circles) the polarization degree shows a similar level of variability, implying that the polarization evolved at least down to $\sim6$\,hrs timescale at some instances during the campaign. The average X-ray polarization is $\approx13\%$, significantly larger than the optical and radio bands. The highest X-ray polarization degree coincides with simultaneous long exposure from \textit{XMM-Newton} and MAGIC. The multi-wavelength evolution during this long exposure is discussed in more details in the next section. \par

In Fig.~\ref{fig:pdeg_ratios}, the temporal evolution of the X-ray-to-optical (black markers) and radio-to-optical (blue markers) polarization degree ratios are shown to illustrate the chromatic behaviour. Besides the polarization being on average higher in the X-rays, a strong variability in the ratios can be seen. On MJD~60291 (December 13 2023, coinciding with the long exposure from \textit{XMM-Newton} and MAGIC), the X-ray polarization degree is above the R-band by a factor of $\approx$ 8-9. For other days (e.g. MJD~60295, December 17 2023), the ratio is close to 1. The increase with energy of the polarization degree is inline with previous observations \citep{2022ApJ...938L...7D, 2024A&A...681A..12K}, but it is the first time that such variations in their respective ratios are reported \citep[see][]{2024A&A...684A.127A}. In fact, in the HSP Mrk~501 (which shares very similar spectral properties with Mrk~421), the ratio between the polarization degrees stayed constant over several IXPE observations \citep{2024A&A...685A.117M}.

The polarization angle from the radio to optical bands shows variations by at most $40^\circ$ over the full observation period (bottom panel in Fig.~\ref{fig:longterm_LC}). The optical polarization angle is consistent with the results from the 2022 campaign \citep{2022ApJ...938L...7D, 2024A&A...684A.127A} and aligns well with the radio jet direction reported in \citet{2022ApJS..260...12W} ( -14.4$^\circ$ $\pm$ 14.2$^\circ$, or 165.6$^\circ$ $\pm$ 14.2$^\circ$ taking into account the $180^\circ$ ambiguity). However, our radio data show systematic shifts between the measured angles at different radio frequencies. It ranges from $\approx 90^\circ$ in the 4.85\,GHz band, and gradually increases with frequency, reaching $\approx 150^\circ$ at 225.5\,GHz, which is much closer to the angle in the optical. This is indicative of a Faraday screen with a rotation measure of $\sim -200$ rad/m$^2$. Alternatively, it may indicate a bending of the jet, which is a relatively common feature in jetted AGNs \citep{1994AJ....108..766B}. The gradual shift towards lower frequencies suggests that the bending occurs in broader regions located downstream of the one dominating the emission beyond optical frequencies, because at the lowest frequencies the radio flux in compact regions is synchrotron self-absorbed \citep{1981ApJ...243..700K}. Further studies are needed to properly assess the origin of this chromatic evolution of the radio polarization angle.\par 

Differently from the other bands, the X-ray polarization angle displays strong, but stochastic-like fluctuations of $\approx100^\circ$, from $\approx 130^\circ$ to $230^\circ$. This behaviour is quite different from the systematic angle rotation of Mrk~421 reported in \citet{2023NatAs...7.1245D} which proceeded over 5 consecutive days with a roughly constant velocity. We stress however that the average value of the polarization angle during December 2023 is $\approx 180^\circ$, aligning well with the results in the optical band and the direction of the radio jet \citep{2022ApJS..260...12W}.

\subsection{Spectral evolution} \label{sec:spectral_evolution}

To investigate the spectral behavior of Mrk~421 during the period studied here and compare it with archival observations, we computed the hardness ratio in the X-ray and VHE regimes using the high- and low-energy flux in each band. Fig.~\ref{fig:HR_xray_vhe} depicts the hardness ratios in the different bands compared to the low- and high-energy flux levels for this campaign. Points with black-filled markers highlight measurement during the IXPE window. We quantified the correlation between the hardness ratio and the fluxes using the Pearson's $R$ coefficient and the corresponding statistical significance in Gaussian units was derived following the prescription of \citet{press2007numerical}. The results are shown in each panel, both for the full 2023 campaign and the IXPE window only. Additionally, for comparison purposes, we plot the observations from the 2022 campaign \citep{2024A&A...684A.127A}, when the first IXPE observations of Mrk~421 were taking place and with the 2017 campaign published in \citet{2021A&A...655A..89M}, which is representative of the typical dynamical flux range of Mrk~421 in the X-rays and at VHE. In comparison to the archival campaigns, the current data lies in a softer regime than usual for the observed high flux level.\par

A strong indication of a harder-when-brighter behavior is observed in the high-energy X-ray and VHE gamma-ray bands over the  full 70-day dataset presented in this manuscript, from November 7 2023 to January 16 2024 (see the right panels in Fig.~\ref{fig:HR_xray_vhe}). The significance of the correlation is $11\sigma$ and $4.4\sigma$ for the 2-10\,keV and $>1$\,TeV bands, respectively. For the low-energy band fluxes, 0.3-2\,keV and 0.2-1\,TeV (see Fig.~\ref{fig:HR_xray_vhe} left panels), however, the significance is much lower and hints towards multiple trends as indicated by the two branches seen in the figures. One follows a steeper trend similar to the one shown in the high-energies. The second one depicts an almost flat relation extending to high fluxes with low hardness ratios. This flatter branch can be associated with the December flare, indicating that no harder-when-brighter behavior is seen during the IXPE time window.\par  

Also in the archival data the harder-when-brighter trend is more visible in the comparison with the high-energy fluxes than for the low-energy band. However, hardness ratios as low as during the IXPE window have not been observed in the 2017 and 2022 campaigns in combination with the high flux levels.\par

For completeness, we present in Appendix~\ref{sec:spectral_fits_xrt} \& Appendix~\ref{sec:spectral_fits_magic} all the spectral parameters for each of the \textit{Swift}-XRT and MAGIC observations during the campaign. For \textit{Swift}-XRT, we show the results of power-law ($dN/dE \propto (E/E_0)^{-\Gamma}$) and log-parabola fits ($dN/dE \propto (E/E_0)^{-\alpha-\beta \log(E/E_0)}$) with a pivot energy $E_0$ fixed to 1\,keV. In the case of MAGIC, the spectra are fitted with a log-parabola model having $\beta$ fixed to 0.5, and a normalization energy set to $E_0=300$\,GeV. The motivation to fix $\beta$ is the following: although we find a significant preference ($>3\sigma$) for a log-parabola model over a simple power law for several of the days, $\beta$ does not exhibit any significant dependence on the flux level. By fixing $\beta$ we still obtain a satisfactory description of all the days, and $\alpha$ can be directly used to evaluate the spectral hardness evolution given that the correlation between $\alpha$ and $\beta$ is removed. The parameters from the MAGIC fits are the intrinsic ones, i.e. after correcting for the effect of extragalactic background light (EBL) absorption using the model of \citet{2011MNRAS.410.2556D}.

\subsection{Long MAGIC/\textit{XMM-Newton} exposure on MJD~60291}

\begin{figure}[t!]
  \centering
  \includegraphics[width=1\columnwidth]{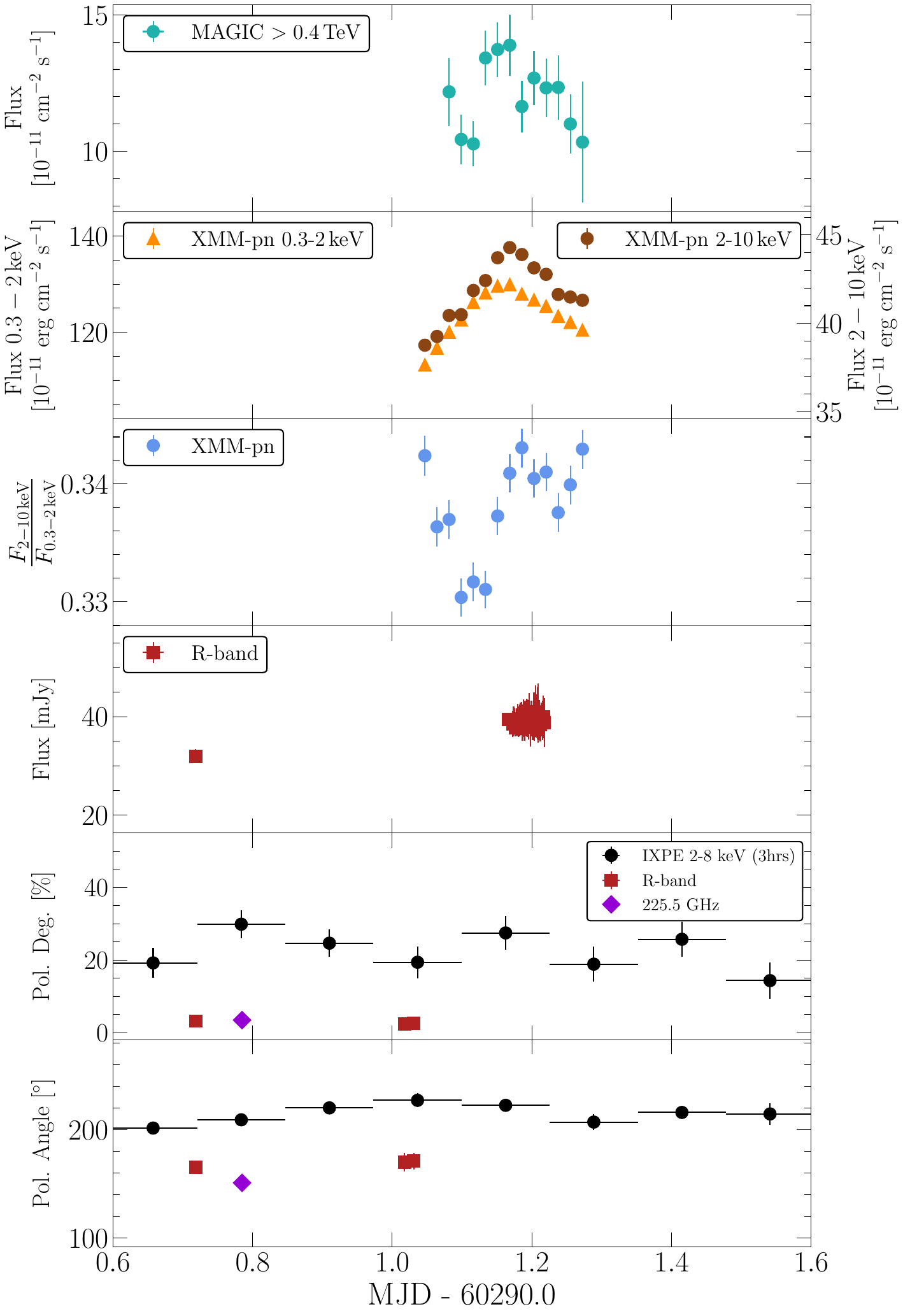}
 \caption{Multi-wavelength light curves during the multi-hour exposure from MAGIC and \textit{XMM-Newton} on MJD~60291 (December 13 2023). The MAGIC fluxes are calculated above 400\,GeV and binned over 25-minutes intervals. The \textit{XMM-Newton} fluxes are computed in the same 25-minutes bins as MAGIC, and in the 0.3-2\,keV and 2-10\,keV bands. In the third panel from the top we show the hardness ratio ($F_{\rm 2-10\,keV}$/$F_{\rm 0.3-2\,keV}$) from the \textit{XMM-Newton} data. The bottom panels show the optical (R-band) fluxes, and the polarization in the radio and optical (R-band) with the same binning as in Fig.~\ref{fig:longterm_LC}. In black markers, we present the X-ray polarization from IXPE binned over 3\,hrs intervals.}
 \label{fig:XMMnight_mwl_lc}
\end{figure}

\begin{figure}[h!]
  \centering
  \includegraphics[width=1\columnwidth]{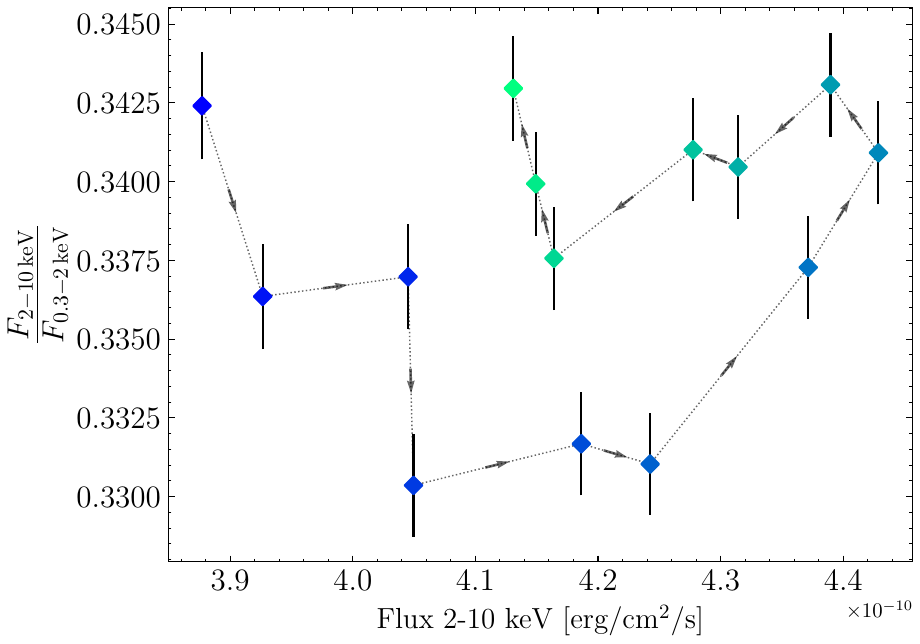}
 \caption{Hardness ratio ($F_{\rm 2-10\,keV}$/$F_{\rm 0.3-2\,keV}$) as function of the 2-10\,keV flux during the \textit{XMM-Newton} observations on December 13 2023 (MJD~60291). The markers are colour-coded with a gradient from dark blue (first time bin) to light green (last time bin), with grey arrows depicting the direction of time, and showing a clear counter-clockwise loop pattern.}
 \label{fig:XMM_hysteresis}
\end{figure}

The organisation of multiband long exposures represents a good opportunity to probe flux and spectral variations down to sub-hour scales, being the timescale over which Mrk~421 (and blazars in general) are known to vary. Such investigations are also essential to provide constraints on the emitting region dimension using causality arguments. Fig.~\ref{fig:XMMnight_mwl_lc} presents the intra-night multi-wavelength light curves on December 13 2023 (MJD~60291) during the simultaneous long exposure with \textit{XMM-Newton} and MAGIC. The total exposure is 16.9\,ks for \textit{XMM-Newton} (in the pn camera) and 15.6\,ks for MAGIC. The MAGIC light curve (top panel) is computed over $20$-min intervals above 0.4\,TeV. The \textit{XMM-Newton} fluxes (second panel from the top) are computed over $\approx15$\,min bins, in the 0.3-2\,keV and 2-10\,keV bands. We also show in the third panel from the top the hardness ratio from the \textit{XMM-Newton} data, which we defined as the ratio between the 2-10\,keV and 0.3-2\,keV fluxes. The lower panels provide the R-band fluxes, multi-wavelength polarization degree and angle, respectively. The IXPE data are binned over 3\,hrs intervals. \par

\begin{figure*}[h!]
    \centering
     \includegraphics[width=\textwidth]{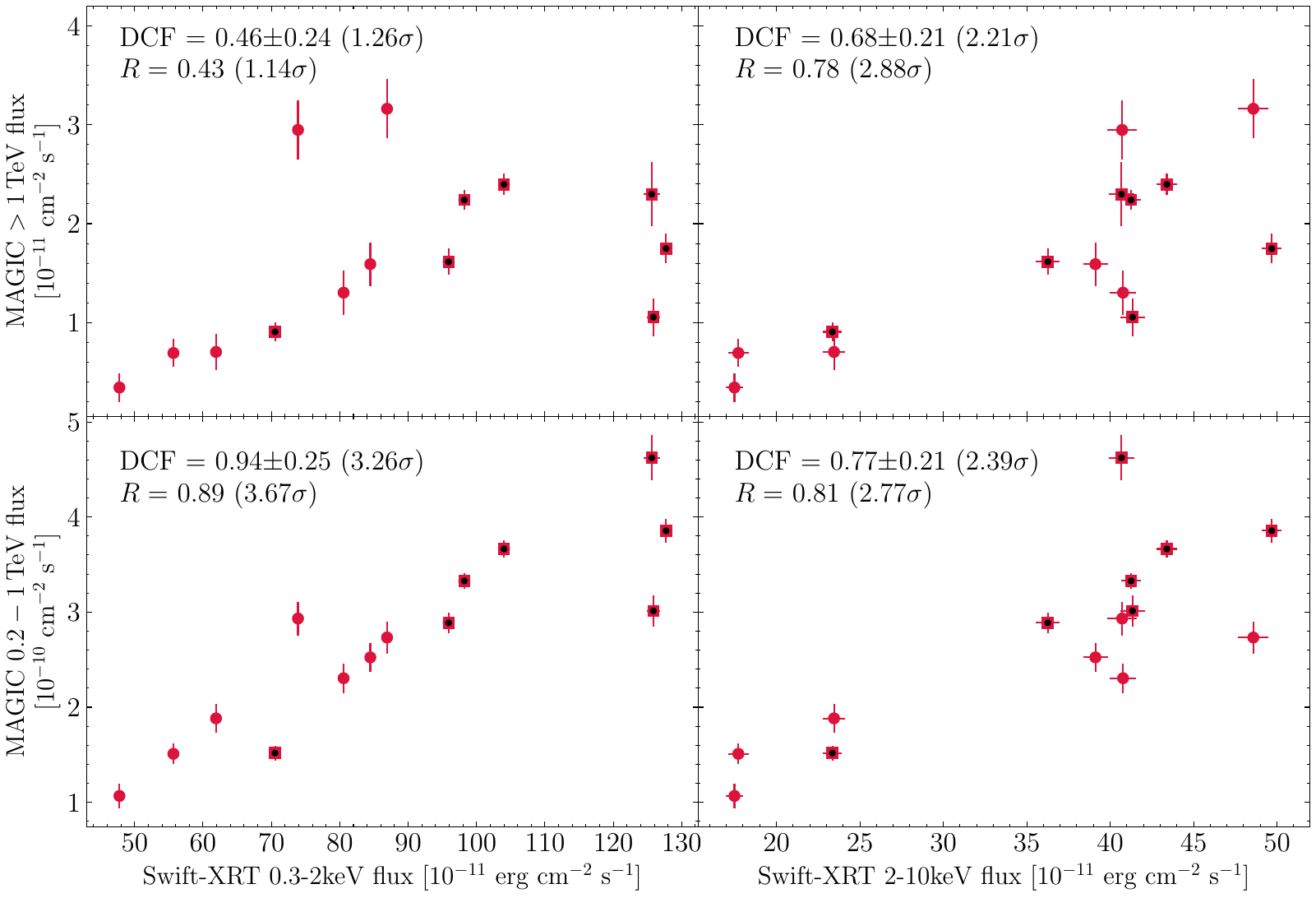}    
    \caption{VHE energy flux versus X-ray flux using MAGIC and \textit{Swift}-XRT observations throughout the multi-wavelength campaign. MAGIC data are nightly binned with a typical exposure time per night of about 40\,min. The \textit{Swift}-XRT observations are binned per observation (typical exposure time of $\approx 15$\,min). Only pairs of measurement within 3\,hours are considered. The MAGIC fluxes are computed in the $>1$\,TeV band (top panels) and in the 0.2-1\,TeV band (lower panels). The \textit{Swift}-XRT fluxes are computed in the 0.3-2\,keV (left panels) and 2-10\,keV bands (right panels). The black filled square markers are the measurements during the IXPE observing window. In each sub-panel, we display the DCF value (and uncertainty) and the corresponding significance in Gaussian $\sigma$ unit in parenthesis. For completeness, the Pearson's $R$ coefficient and its significance are also given. The significance is determined based on dedicated Monte-Carlo simulations (see Sect.~\ref{sec:MWL_correlation} for more details).} 
    \label{fig:xrt_vs_magic}
\end{figure*}

The MAGIC data show a non-significant ($<3\sigma$) flux variability. By fitting the data with a constant model, we obtain $\chi^2/{\rm dof}=17.2/11$, implying that the hypothesis of a non-variable emission is rejected at a significance of only $1.3 \sigma$. Differently, the \textit{XMM-Newton} light curves reveal significant variability, although with a moderate flux amplitude at the level of 10-15\% in both bands. The hardness ratio also varies significantly, implying sub-hour spectral variability in the X-rays. From a constant fit, we obtain $\chi^2/{\rm dof}=96.5/13$, and the hypothesis of a constant spectral behaviour is rejected at a significance of $7.8 \sigma$. The hardness ratio is lower during the rising phase of the flux than during the decaying phase, hinting towards a delay of the high-energy flux compared to the low-energy flux. In fact, as it can be seen in the light curve, the 2-10\,keV band reach its maximum about 15\,min after the 0.3-2\,keV band.

Fig.~\ref{fig:XMM_hysteresis} presents the hardness ratio plotted versus the 2-10\,keV flux. Grey arrows give the direction of time. The data show an evident loop in counter-clockwise direction. This hysteresis pattern is indicative of a delay of the higher energy flux with respect to the lower energy one. As discussed in \citet{1998A&A...333..452K} and in Sect.~\ref{sec:discussion}, such a lag is likely the manifestation of a regime where the radiation cooling timescale is comparable to the particle acceleration timescale. \par

In Fig.~\ref{fig:XMMnight_mwl_lc}, the IXPE polarization measurements are binned over 3\,hrs intervals to search for potential short timescale variability contemporaneous to the long MAGIC/\textit{XMM-Newton} exposure. The polarization degree is not varying significantly, and is stable around a value of 20-25$\%$. This is a factor of $\approx$ 8-9 higher than in the optical band. As for polarization angle, some slow variations in the order of $20^\circ$ are visible. The average X-ray polarization angle exhibits an offset by $\approx$ 60$^{\circ}$-70$^{\circ}$ with respect to the optical and radio polarization angles (which are within $\approx10^{\circ}$ from each other).

\section{Correlated variability} \label{sec:MWL_correlation}

This section first reports on the quantification of the intra-band flux correlations, focusing on the VHE and X-rays. Owing to the unprecedented variability seen in the IXPE data, we also searched for potential correlation of the X-ray polarization properties with the flux and the behaviour in other bands.

\subsection{Intra-band flux correlation}
We investigated the correlation between the VHE and X-ray fluxes in light of their strong variability throughout the campaign, as well as their expected correlated variability within leptonic models \citep{2007A&A...462...29G, 2008ApJ...677..906F}. To do so, we correlated the MAGIC and \textit{Swift}-XRT fluxes, considering all possible combinations between the respective sub-energy bands, i.e. 0.2-1\,TeV \& $>1$\,TeV for MAGIC, and 0.3-2\,keV \& 2-10\,keV for \textit{Swift}-XRT (see Fig.~\ref{fig:longterm_LC}). We matched pairs of measurements that are within a time window of less than 3\,hrs from each other, that is well below the flux doubling/halving timescales reported at those energies during this campaign. The correlation was then quantified using the discrete correlation coefficient \citep[DCF;][]{1988ApJ...333..646E} assuming a time-lag $t_{\rm lag}=0$ between the light curves. For completeness and comparison purposes, we also show in each panel the widely-used Pearson correlation coefficient $R$. We note the DCF is generally preferred over the Pearson correlation coefficient when dealing with two uneven time series. The DCF method also has the advantage to naturally incorporate measurement uncertainties \citep{1988ApJ...333..646E}.\par 

The resulting above-mentioned flux-flux plots are shown in Fig.~\ref{fig:xrt_vs_magic}. For clarity, black filled markers correspond to observations during the IXPE observing window. In each panel, the DCF and $R$ values are provided together with their significance (in Gaussian $\sigma$ units). To assess the significance of the correlation, we followed a procedure described in \citet{2024A&A...684A.127A} that is based on the code described in \citet{axel-thesis}. In summary, we simulated $10^5$ uncorrelated light curves for each energy bands that have the same sampling and binning as the observations. The light curves were simulated assuming a power spectral density (PSD) that follows a power law with index fixed to $-1.4$ for all energy bands. Such a PSD index agrees with recent works on Mrk~421 \citep{2010_421_paper, 2015A&A...576A.126A, 2015ApJ...798...27I}. The simulated light curves preserve the probability density function of the data using the method from \citet{2013MNRAS.433..907E}. Finally, the significance was estimated by comparing the observed DCF and $R$ values with the distribution from the simulated light curves.\par 

A positive VHE/X-ray correlation is typical in Mrk~421, and has been reported multiple times both during low \citep[see e.g.,][]{2015A&A...576A.126A,2016ApJ...819..156B,2021MNRAS.504.1427A} and high activity \citep[see e.g.][]{2015A&A...578A..22A,2020ApJS..248...29A}, suggesting a common underlying particle population. During some flaring states, a tighter correlation of the X-rays with the $\gtrsim1$\,TeV band than with the 0.2-1\,TeV band \citep[see e.g.,][]{2020ApJS..248...29A} has been found. However, this does not happen for all flares \citep[see e.g.,][]{2021A&A...655A..89M}. Interestingly, during the 70-day campaign in 2023-2024 considered in this study, there is substantial variability in both X-rays and VHE gamma rays, but the correlation between the various bands in X-rays and VHE gamma rays is only marginally significant (see inlaid information in the panels from Fig.~\ref{fig:xrt_vs_magic}). The magnitude of the DCF and Pearson correlation is high ($>$0.7) for most of the bands, but the significance of these values is not large due to the large scattering in the data points. There is an apparent overall correlation pattern in all the panels. Nonetheless, there are some groups of data points that are far away from the overall pattern. For instance, within the IXPE window (see black filled markers), one can see a few observations where the $>1$\,TeV (0.2-1\,TeV) flux varied by more than a factor 2 (1.5) for a very similar X-ray flux state. Such deviations from the overall trend, which indicate different emission mechanisms at work, increase the calculated error for the DCF value, and naturally also decrease the significance of the correlations.

\subsection{Polarization}
We searched for possible correlations between the X-ray polarization degree and angle, as well as between the X-ray polarization properties and flux. We also correlated the X-ray polarization with the VHE flux (0.2-1\,TeV). The latter investigation is motivated by the fact that IXPE probes the 2-8\,keV band that is originating from electrons with similar energies as the one that are also responsible for the VHE emission within leptonic models \citep{2007A&A...462...29G, 2008ApJ...677..906F}. The correlation was conducted using the IXPE data binned into 6\,hrs and 12\,hrs intervals as presented in Fig.~\ref{fig:longterm_LC}. No significant ($>3\sigma$) evidence for an underlying correlation pattern is found, neither with any of the X-ray and VHE bands (see Appendix~\ref{sec:xray_pol_corr}, Fig.~\ref{fig:pdeg_vs_fluxes} and Fig.~\ref{fig:pa_vs_fluxes}), nor between the polarization degree and angle themselves (see Appendix~\ref{sec:xray_pol_corr}, Fig.~\ref{fig:pdeg_vs_pa_ixpe}). The X-ray polarization (degree and angle) also does not show a significant correlation with the X-ray or VHE hardness ratios (see bottom panels in Fig.~\ref{fig:pdeg_vs_fluxes} and Fig.~\ref{fig:pa_vs_fluxes}). In summary, the X-ray polarization behaves erratically, and the variations do not seem to be related to the flux level nor the spectral shape.\par

\begin{table*}[h!]
\caption{\label{tab:ssc_parameters_fixed} Fixed parameters of the two-component leptonic scenario used to model the broadband SEDs}
\centering
\begin{tabular}{l c @{\hskip 0.3in} c c}     
\hline\hline
Parameters   & ``extended'' zone &  ``compact'' zone \\
\hline
\hline
$\delta$ & 60 & 60 \\
$R'$ [$10^{16}$cm] & 1.5 & * \\
$p$ & 2.2 & * \\
$\gamma'_{min}$ & $2\times10^{3}$ & $2\times10^{4}$  \\
$\gamma'_{\rm cutoff}$ & $2.7\times10^{4}$ & $1.3\times10^{5}$  \\ 
\\\hline 
\end{tabular}
\tablefoot{ We refer the reader to Section~\ref{sec:modelling} for a detailed description of each parameter. The parameters for the ``compact'' zone denoted with a ``*'' are evolving daily, and can be retrieved in Table~\ref{tab:ssc_parameters_evolving}.  }
\end{table*}
 \begin{table*}[h!]
\caption{\label{tab:ssc_parameters_evolving} Evolving parameters of the theoretical leptonic scenario used to model the broadband SEDs.}
\centering
\begin{tabular}{l @{\hskip 0.7in} c c c @{\hskip 0.3in} c c c c}     
\hline\hline
 & \multicolumn{2}{c}{``extended'' zone} & & \multicolumn{4}{c}{``compact'' zone} \\
Day  & $B'$ & $N'_e$ & & $^\star R'$ &  $B'$ & $N'_e$ & $p$ \\
\lbrack MJD \rbrack  & [$10^{-2}$G] & [$10^{-2}$ cm$^{-3}$] & & [$10^{16}$cm] & [$10^{-2}$G] & [$10^{-2}$ cm$^{-3}$] &   \\
\hline
\hline
60287 & 2.3 & 37.7 & & 0.95 & 4.0 & 3.5 & 2.50 \\
60288 & 2.3 & 53.8 & & 0.80 & 5.5 & 2.36& 2.50 \\
60290 & 2.3 & 53.8 & & 0.50 & 5.5 & 8.5 & 2.15 \\
60291 & 2.3 & 56.5 & & 0.40 & 8.0 & 13.2 & 2.50 \\
60292 & 2.3 & 45.7 & & 0.96 & 4.0 & 3.1 & 2.30 \\
60294 & 2.3 & 48.5 & & 0.90 & 5.5 & 2.8 & 2.50 \\
60295 & 3.5 & 22.9 & & 1.32 & 5.0 & 1.4 & 2.50 \\
60296 & 3.0 & 37.7 & & 0.81 & 7.5 & 2.7 & 2.60 \\
60299 & 3.5 & 25.3 & & 1.23 & 4.0 & 2.1 & 2.30 \\

\hline 
\end{tabular}
\tablefoot{We mark the parameter $R'$ with a $^\star$ to stress that this is not a free parameter of the model as it is determined from the measured optical-to-X-ray polarization degree. We refer the reader to Section~\ref{sec:modelling} for a detailed description of each parameter. }
\end{table*}

\section{Theoretical description of the flare evolution using constrains from the X-ray polarization measurements with IXPE} 

\label{sec:modelling}

We modelled the broadband evolution during the IXPE observing window assuming a leptonic scenario \citep[][]{1992ApJ...397L...5M, 1996ASPC..110..436G, Tavecchio_Constraints}. Leptonic scenarios assume that the radio to X-ray originates from electron-synchrotron radiation, while the gamma-ray photons are produced by electron inverse-Compton scattering off the synchrotron photons. Such models are supported by the correlation X-ray/VHE that has been observed in the emission of Mrk~421, even if the period considered here led to only marginally significant correlations (despite the relatively large magnitude in the correlation), as reported in Fig.~\ref{fig:xrt_vs_magic}. For this, we first computed daily SEDs from radio to VHE around each day that includes a MAGIC observation. The MAGIC SEDs were extracted after correcting for the EBL absorption using the model of \citet{2011MNRAS.410.2556D}. We complemented the MAGIC SEDs with strictly simultaneous X-ray data from \textit{Swift}-XRT and also from \textit{XMM-Newton} during the long exposure on December 13 2023 (MJD~60291). The \textit{Fermi}-LAT observations were averaged over 3-day intervals centered at the MAGIC observing time because of the limited sensitivity to resolve the spectrum of Mrk~421 below daily timescales and the low variability in this waveband. We included R-band and UV observations that are the closest in time to MAGIC. Due to the sparse sampling of \textit{Swift}-UVOT (Fig.~\ref{fig:longterm_LC}), some UV observations have a time difference reaching a maximum of 8.5\,days with respect to the MAGIC one. Due to the moderate variability in those bands (see Fig.~\ref{fig:longterm_LC} and Fig.~\ref{fig:fvar}), they can nevertheless be considered as a good proxy for the UV emission during the X-ray and VHE gamma-ray measurements. Finally, we added contemporaneous radio data (simultaneous on timescales of several days) for completeness, they are not included in our model given that they may receive significant contribution from multiple regions (not considered in our model for simplicity) further downstream the jet. In total, this provides us with 9 SEDs (between MJD~60287 and MJD~60299, December 9 2023 to December 21 2023), which we modelled within the scenario described below. We used the same code as in \citet{2021A&A...655A..89M, 2024A&A...685A.117M}, which employs routines from the \texttt{naima} package \citep{2015ICRC...34..922Z} to compute the synchrotron and inverse-Compton emissivities.\par 

\begin{figure*}[h!]
    \centering
    \begin{subfigure}[b]{0.49\textwidth}
        \centering
        \includegraphics[width=\textwidth]{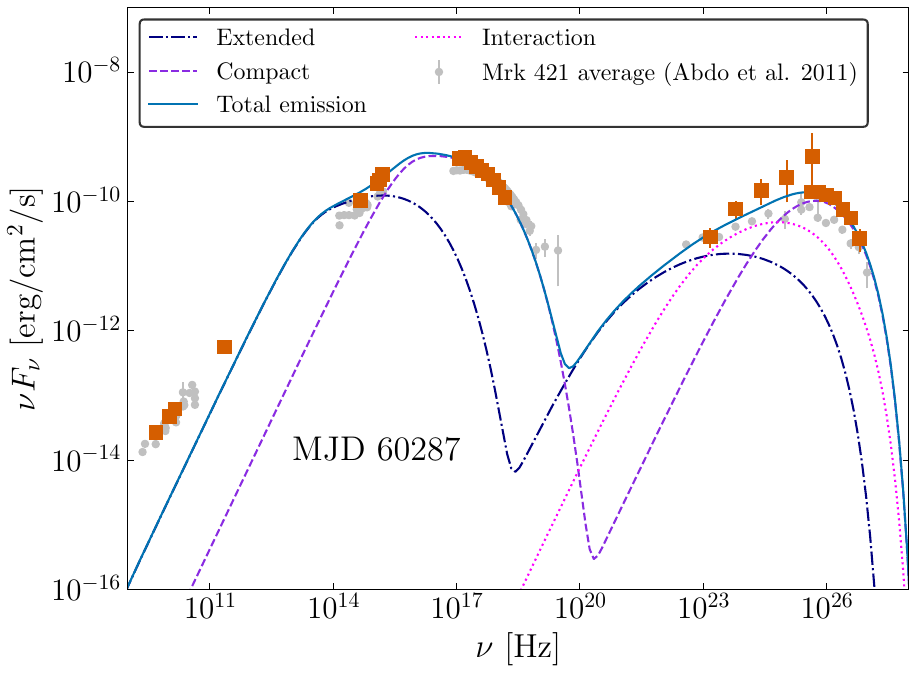}
    \end{subfigure}
    \begin{subfigure}[b]{0.49\textwidth}  
        \centering 
        \includegraphics[width=\textwidth]{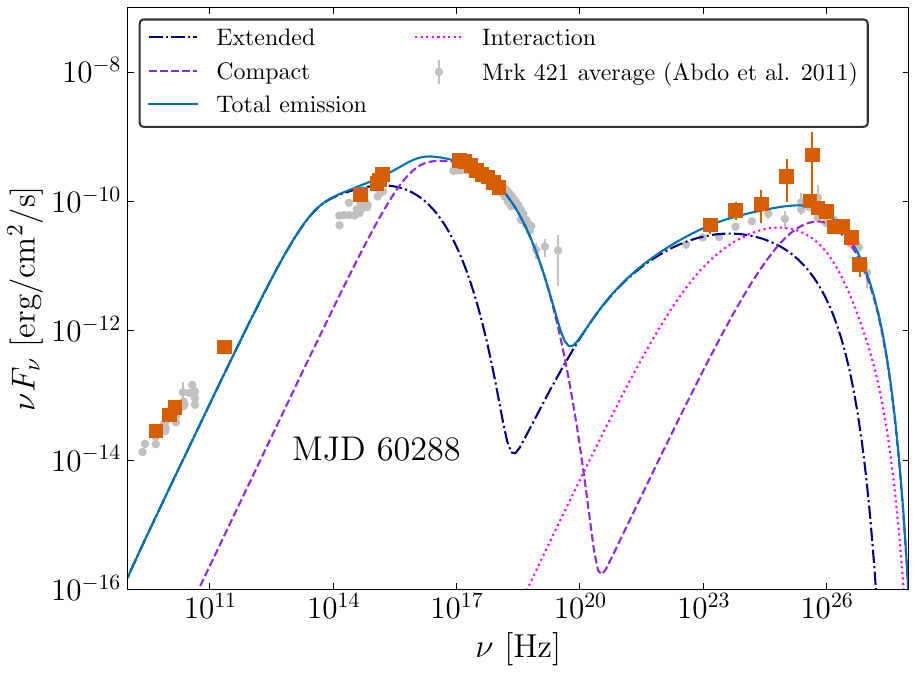}
    \end{subfigure}
    \vspace{0.1in}
    \begin{subfigure}[b]{0.49\textwidth}   
        \centering 
        \includegraphics[width=\textwidth]{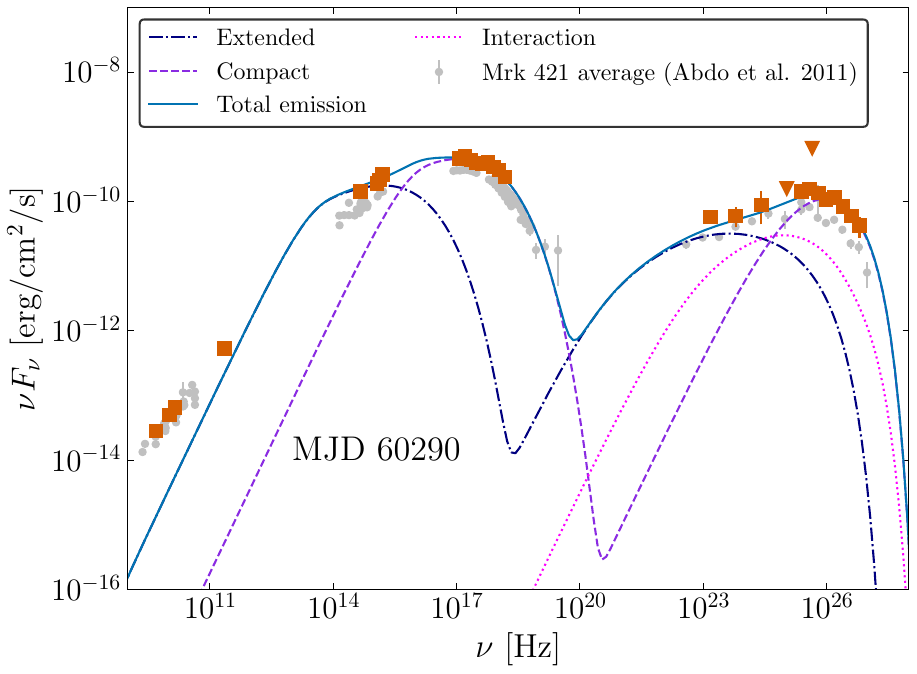}
    \end{subfigure}
    \begin{subfigure}[b]{0.49\textwidth}   
        \centering 
        \includegraphics[width=\textwidth]{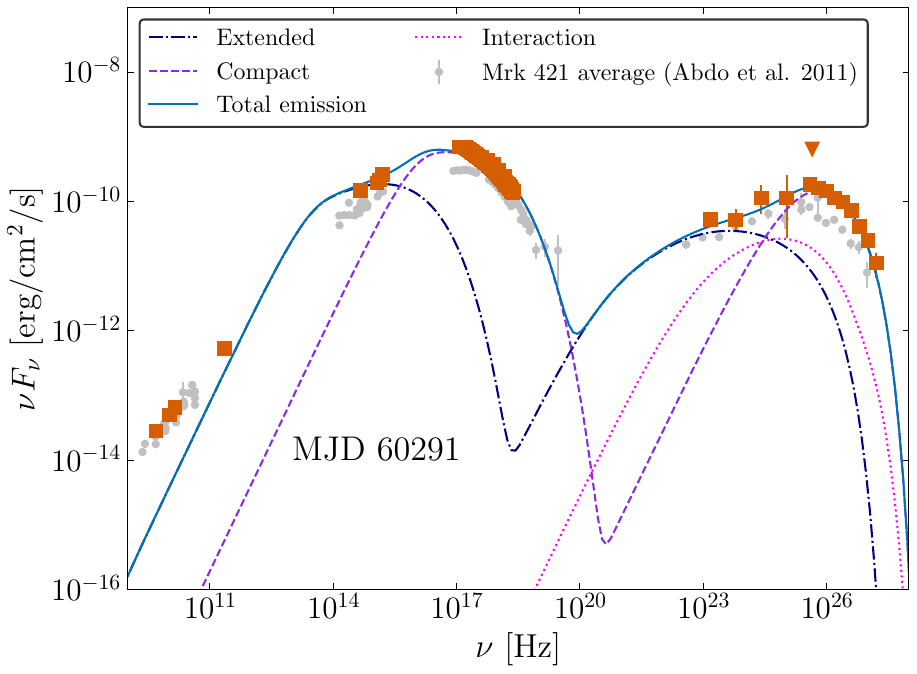}
    \end{subfigure}
    \begin{subfigure}[b]{0.49\textwidth}
        \centering
        \includegraphics[width=\textwidth]{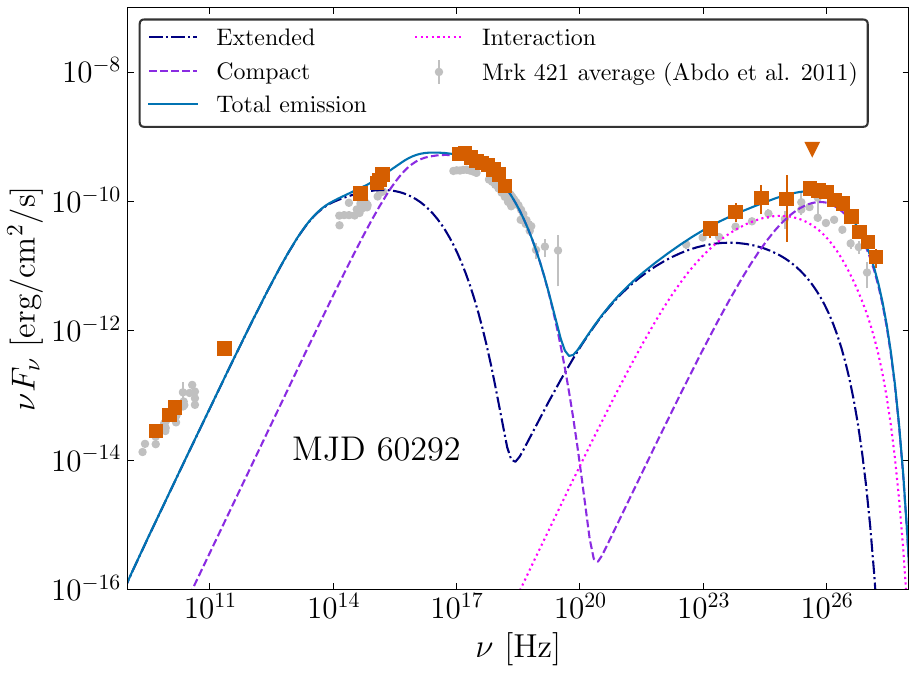}
    \end{subfigure}
    \begin{subfigure}[b]{0.49\textwidth}  
        \centering 
        \includegraphics[width=\textwidth]{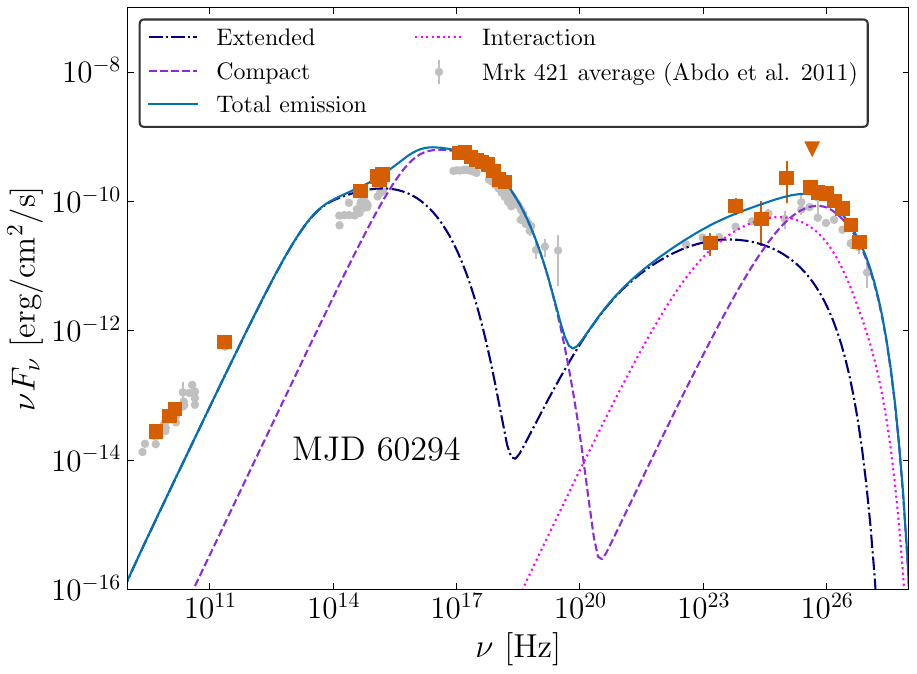}
    \end{subfigure}  
    \caption{Theoretical models for of the broadband SEDs during the IXPE observing window, from MJD~60287 to MJD~60299 (December 9 2023 to December 21 2023). The observations, spanning from the radio to VHE, are shown by the orange markers. The dashed violet curve is the emission originating from the ``compact'' zone, located nearby the shock front. The dot-dashed blue line is the contribution from the ``extended'' zone, which spans a larger volume downstream the shock. The emission produced by the interaction between the two zones is plotted in a pink dotted curve. The solid blue line shows the sum of all components. For comparison purposes, we plot with grey markers the average state of Mrk~421 \citep[taken from][]{abdo:2011}. The values for the model parameters can be found in Table~\ref{tab:ssc_parameters_fixed} and Table~\ref{tab:ssc_parameters_evolving}. We refer the reader to Sect.~\ref{sec:modelling} for more details on the modelling procedure.\\ } 
    \label{fig:sed_modelling}
\end{figure*}
\begin{figure*}[h!]
    \centering
    \begin{subfigure}[b]{0.49\textwidth}
        \centering
        \includegraphics[width=\textwidth]{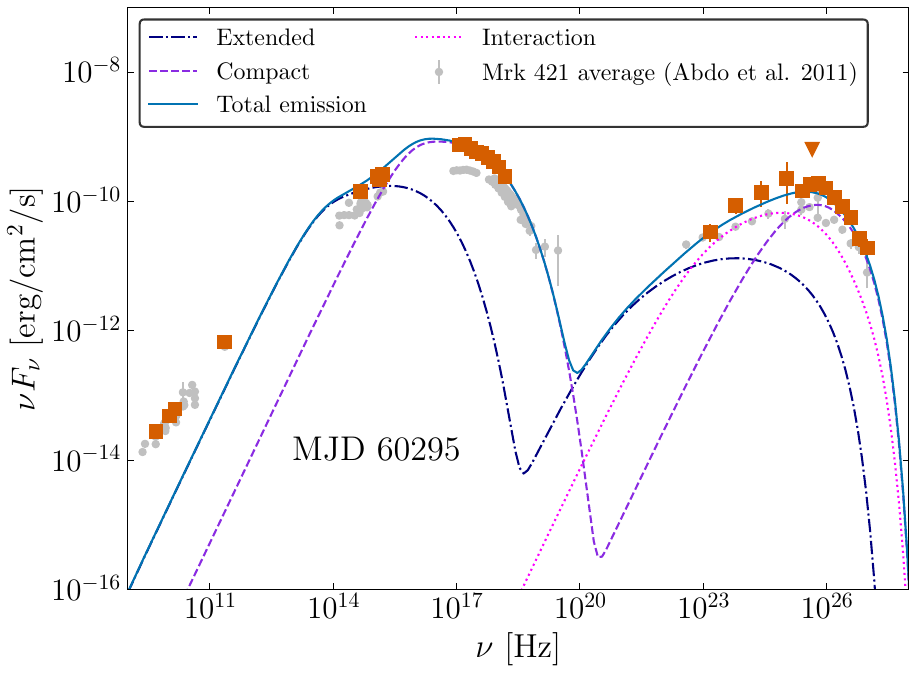}
    \end{subfigure}
    \begin{subfigure}[b]{0.49\textwidth}  
        \centering 
        \includegraphics[width=\textwidth]{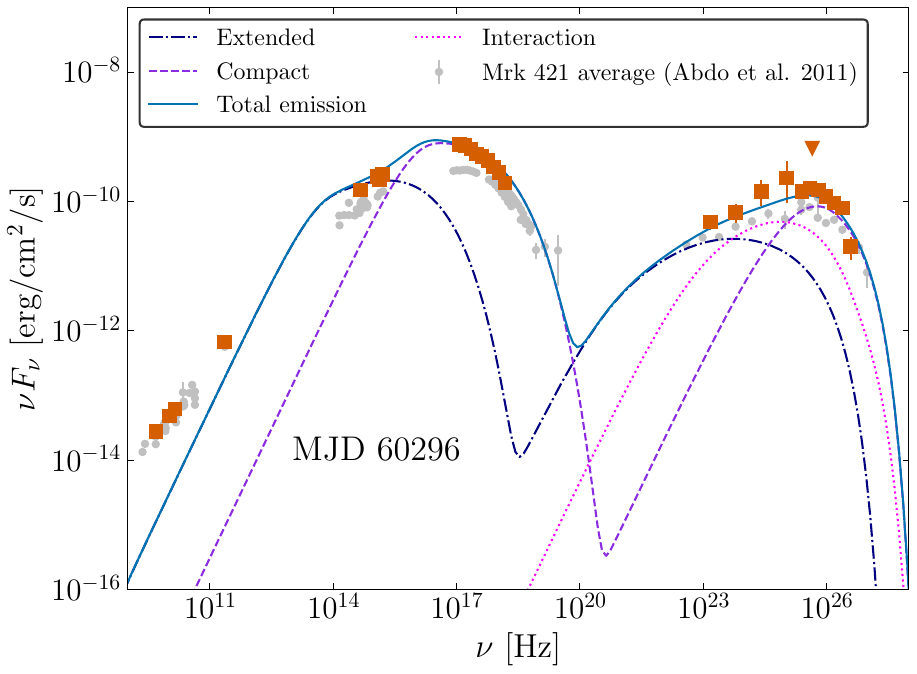}
    \end{subfigure}
    \begin{subfigure}[b]{0.49\textwidth}  
        \centering 
        \includegraphics[width=\textwidth]{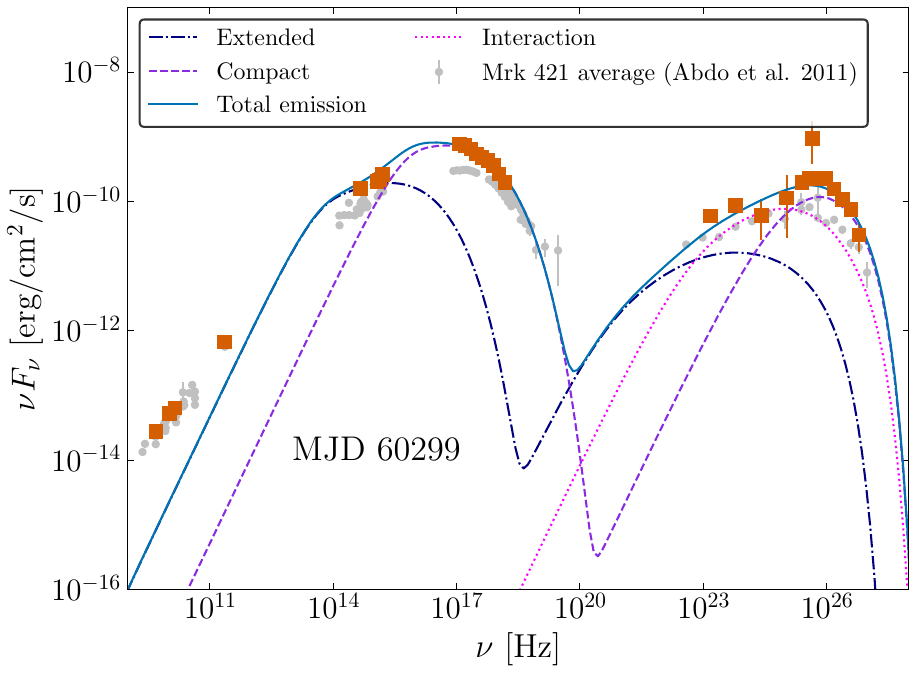}
    \end{subfigure}
    \caption{Same as Fig.~\ref{fig:sed_modelling} for the last three nights during the IXPE observing window.} 
    \label{fig:sed_modelling2}
\end{figure*}

Our model aims at qualitatively capturing the radio-to-X-ray polarization properties that suggest an energy stratification of the emitting region (see Sect.~\ref{sec:pol_behavior}). In a similar approach to \citet{2024A&A...685A.117M}, we assumed a morphology consisting of two overlapping, spherical zones: a ``compact'' zone near the acceleration site, and an ``extended'' zone that occupies a larger volume downstream the jet. The ``compact'' zone contains freshly accelerated and energetic electrons that dominate the X-ray and $>100$\,GeV emission. The ``extended'' zone, populated by less energetic electrons, is dominant in the UV/optical and <100\,GeV bands. In line with \citet{2022Natur.611..677L, 2016MNRAS.463.3365A, 2014ApJ...780...87M}, the observed higher polarization degree in the X-ray band compared to the radio/optical is due to the confinement of the freshly accelerated electrons in a smaller region near the shock front (mimicked by the ``compact'' zone), which compresses the magnetic field leading to a more ordered magnetic field structure. When electrons subsequently cool and advect towards larger regions (mimicked here by the ``extended'' zone) in which the degree of magnetic field turbulence increases significantly, a drop of the polarization degree at lower frequencies is expected. Given the very different polarization variability patterns observed in the optical and X-ray bands, it is required that the ``compact'' zone remains subdominant in the UV/optical compared to the ``extended'' zone. Finally, since we assumed that the two regions are overlapping, our code also includes the emission resulting from the interaction of the two zones since the (synchrotron) emission from the two regions provide an additional target photon field for each other for the inverse-Compton scattering. The up-scattering of the ``extended'' zone field by the electrons in the ``compact'' zone leads to an inverse-Compton luminosity comparable to the one resulting from the up-scattering of the ``compact'' zone field by the ``extended'' zone electrons. The only difference is that the two components are slight shifted in frequency due to the different electron and photon energies populating each zone. \par 

Our assumption of using two distinct components is naturally a simplification of the reality because one would rather expect a continuum of several regions contributing to the different parts of the SED. A detailed treatment of the turbulences, electron advection and diffusion is beyond the scope of this work. \par

We exploit the polarization data to constrain the relative size between the ``compact'' and ``extended'' zones for the 9 SEDs modelled here. We assume that each of the region is made of $N$ turbulent plasma cells of identical magnetic field strength, but with random orientation. In such a configuration, the expected average polarization degree from each zone can be approximated as $P_\text{deg} \approx 75\%/\sqrt{N}$ \citep{2014ApJ...780...87M, 2021Galax...9...37T}. The measured ratio between the X-ray and optical polarization degree ($P_\text{X-ray,deg}/P_\text{opt,deg}$; see Fig.~\ref{fig:pdeg_ratios}) was then used to estimate the relative difference, $l$, between the number of turbulent cells in the ``extended'' zone and the ``compact'' zone. Under the assumption that cells in both zones roughly span the same spherical volume, one derives a relative difference in radius of $l^{1/3}$ between the two regions. Here, we fixed the radius of the ``extended'' zone to $1.5\times10^{16}$\,cm (in line with constraints from the light crossing time), and then determined the radius of the ``compact'' zone using the observed ratio $P_\text{X-ray,deg}/P_\text{opt,deg}$ during each day. The choice of fixing the radius of the ``extended'' zone is motivated by the low (flux and polarization) variability in the UV/optical bands.\par

In order to limit the degrees of freedom of the used theoretical model, we further made the following assumptions (primed quantities are expressed in the source reference frame):

\begin{itemize}
    \item The Doppler factor $\delta$ was fixed to 60 in the ``compact'' zone. Such a Doppler factor is somewhat larger than those typically adopted in the literature for Mrk~421, which are in the range $\sim$20-40 \citep{abdo:2011, 2016ApJ...819..156B, 2019MNRAS.487..845B}, although $\delta\approx60$ is still consistent with the modelling of the March 2008 flare presented in \citet{2012A&A...542A.100A}. The rather soft X-ray spectrum gives rise to a large separation between the two SED peak frequencies, hence requiring a high Doppler factor within a leptonic scenario. Using values significantly lower than 60 (e.g. by a factor two) leads to VHE spectra softer than the observed ones, and hence we excluded them. To illustrate this, Appendix~\ref{sec:modelling_30_test} reports a modelling of the broadband SED from MJD~60291 (December 13 2023), which corresponds to the day with simultaneous MAGIC/\textit{XMM-Newton} observations, using $\delta=30$.\par 
    We would like to point out that $\delta$ values below 60 for the ``extended'' zone would describe satisfactorily the optical and \textit{Fermi}-LAT data in the part of the SED dominated by the ``extended'' zone. However, for the sake of simplicity, and to limit the number of free parameters, we set $\delta=60$ in both zones. We note that the radius of the two zones differs by less than a factor of four, and that the usage of very different Doppler factors between the two zones would imply a strong deceleration of the flow on spatial scales comparable to the size of the emission, that is $\sim10^{16}$\,cm ($\sim10^{-2}$\,pc). This is significantly smaller than what is observed in FR I galaxies \citep{2005MNRAS.358..843H}. \\
    
    \item In both regions, the electron distribution follows a power-law distribution with exponential cutoff: $dn/d\gamma' \propto \gamma'^{-p} \, \exp{\left( -\gamma'/\gamma'_{\rm cutoff} \right)}$, for $\gamma'_{\rm min} < \gamma' < \gamma'_{\rm max}= +\infty$. In the ``compact'' zone, the slope can be well constrained by the X-ray data. For the ``extended'' zone, $p$ was fixed to the canonical value of 2.2 for relativistic shocks \citep{2000ApJ...542..235K} given that the UV/optical and MeV/GeV coverage does not provide a strong constrain on this parameter. \\

    \item In the ``compact'' zone, we fixed $\gamma'_{\rm min} = 2 \times 10^{4}$ and $\gamma'_{\rm cutoff} = 1.3 \times 10^{5}$. Given the magnetic field strength necessary to adequately reproduce the SEDs ($\approx$0.05-0.08\,G), such a large $\gamma'_{\rm min}$ is needed so that the synchrotron flux from the ``compact'' zone (responsible for the X-rays) becomes subdominant at lower energies in the UV/optical. $\gamma'_{\rm cutoff}$ was selected to provide a good description of the X-ray and VHE spectral shapes at the highest energies for all days. \\
    
    \item The electron distribution in the ``extended'' zone is defined with $\gamma'_{\rm min} = 2 \times 10^{3}$ and $\gamma'_{\rm cutoff} = 2.7 \times 10^{4}$. $\gamma'_{\rm min}$ is limited from below in order to not overshoot the radio data in the 225.5\,GHz band, while $\gamma'_{\rm cutoff}$ is set such that the ``extended'' zone does not dominate the overall emission beyond the UV band.
    
\end{itemize}

Based on those assumptions, the only remaining free parameter in our model for the ``extended'' zone is the magnetic field $B'$ and the electron density $N'_e$. In the ``compact'' zone the remaining free parameters are $B'$, $N'_e$ and $p$, which we varied for each day in order to properly describe the X-ray and VHE data. The results are shown in Fig.~\ref{fig:sed_modelling} and Fig.~\ref{fig:sed_modelling2}. The observations are plotted with orange markers. The contribution from the ``extended'' and ``compact'' zones are shown in blue and violet solid lines, respectively. The inverse-Compton emission resulting from the interaction between the two zones is plotted with a pink dotted line and the sum of all components is depicted with a solid light blue curve. For all days a good description of the observations is achieved. We summarize in Table~\ref{tab:ssc_parameters_fixed} the parameters of the two emitting zones that are fixed, and in Table~\ref{tab:ssc_parameters_evolving} the evolving parameters in the ``compact'' and the ``extended'' component.\par

The model does not evolve the particle population self-consistently by taking into account the radiation cooling. Nonetheless, according to the $B'$ strength required to well reproduce the SED and assuming that the particles escape the regions on timescales $t'_{\rm esc} \approx R'/c$, the synchrotron cooling break is above $\gamma'_{\rm cutoff}$ by more than a factor of a few (for both zones). The system is therefore in a slow-cooling regime, i.e. the particles likely escape the emitting region prior to any significant cooling. The cooling effects are thus neglected when modelling the SED evolution.\par

\begin{figure}[h!]
  \centering
  \includegraphics[width=1\columnwidth]{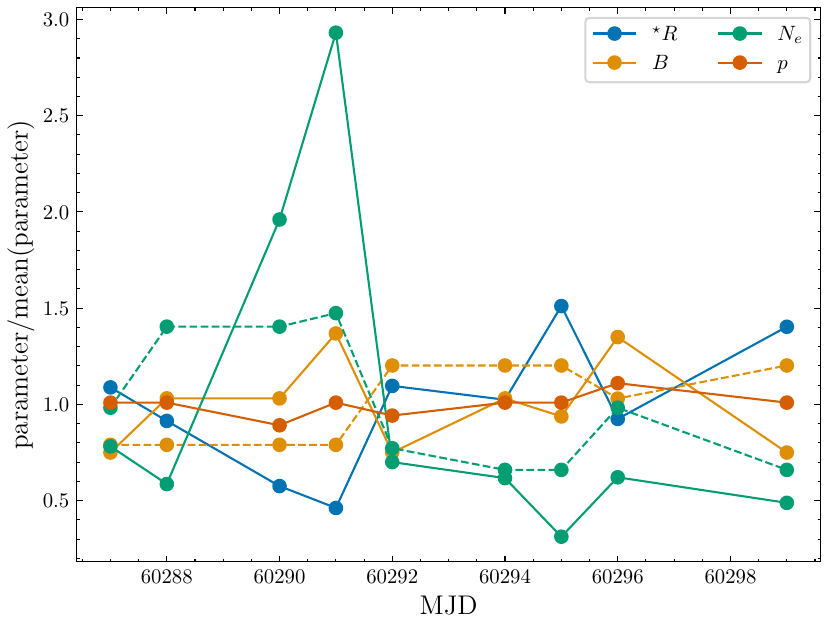}
 \caption{Relative evolution of the evolving parameters in the ``compact'' zone (solid lines) and ``extended'' zone (dashed line). Each parameter, plotted with a different color, is normalized to its average value.}
 \label{fig:parameter_evol}
\end{figure}

Fig.~\ref{fig:parameter_evol} shows the relative evolution for each of the parameters in the ``compact'' zone (solid lines) and in the ``extended'' zone (dashed lines). The radius, $R'$, and the electrons density within the zone, $N'_e$, are the two parameters that vary the most. The radius, determined from the optical-to-X-ray polarization degree varies up to a factor 2 along the flare, and the density evolves by a factor 3 at least. The magnetic field, $B'$, changes moderately, by less than 40\%. Regarding the slope of the electrons $p$, it varies by less than 10\%, which is expected given the moderate spectral variability reported in Sect.~\ref{sec:spectral_evolution}. We provide in Sect.~\ref{sec:discussion} a physical interpretation and a more detailed discussion of the proposed model. \par

\section{Discussion and summary} \label{sec:discussion}

In this work, we have studied, for the first time for a HSP blazar, a bright state of Mrk~421 from radio-to-VHE with simultaneous X-ray polarization measurements using IXPE. The 0.2-1\,TeV flux reached a peak activity slightly above 2\,C.U., which is more than a factor 4 higher than the average state at those energies \citep{2014APh....54....1A}. A bright X-ray flux is also detected in the 0.3-2\,keV band, and is comparable to the level recorded during archival flaring states, such as the one in March 2010 \citep{2015A&A...578A..22A}. Until now,  multi-wavelength studies including X-ray polarization have only probed HSPs in low or quiescent states \citep[see e.g.][]{2024A&A...684A.127A, 2024A&A...685A.117M}. Therefore, the study presented with data from this campaign offers the most complete characterization of a HSP blazar during a high activity so far, hence providing crucial insights into the physical origin of blazar flares.\par 

\subsection{Interpretation of the multi-wavelength polarization variability during the VHE flare}

The polarization degree shows a significant chromatic behavior, indicating an energy-stratification of the emitting zone. The average X-ray polarization degree is about 3 times higher than the R-band, and about 7 times higher than in the radio. Such an energy dependence is comparable to previous results on Mrk~421, and also for the HSP Mrk~501 \citep{2022Natur.611..677L, 2022ApJ...938L...7D}. However, the drastic variations (by a factor $\sim8$) of the X-ray polarization degree  happening down to $\sim6$\,hrs timescales is a novel trend not yet observed in earlier observations. In addition, the X-ray polarization angle varies significantly by about $100^\circ$ on day timescales. These variations of the X-ray polarization do not reveal any evident structured pattern and are consistent with a stochastic trend. This differs from the steady rotation seen in 2022 \citep{2023NatAs...7.1245D} that occurred over about 5 days with a roughly steady angular velocity and constant polarization degree. We stress however that the average X-ray polarization angle during December 2023 aligns closely with the one in the optical and with the radio jet axis \citep{2022ApJS..260...12W}. This suggests a partial ordering of the magnetic field perpendicular to the jet, possibly caused by the presence of a shock compressing the helical field structure \citep{2017Galax...5...63M, 2018MNRAS.480.2872T}. The stochastic nature of the polarization angle and degree evolution is then likely due to the plasma being highly turbulent before crossing the shock front \citep{2014ApJ...780...87M}. We note that magnetic reconnection is also expected to generate rapid and strong variability of the polarization \citep{2022ApJ...924...90Z},  but the polarization angle is expected to reach any orientation with respect to the jet. However, the average angle measured with IXPE remains closely aligned with the jet and the optical band, suggesting that magnetic reconnection does not dominate the acceleration of the electrons responsible for the X-ray and VHE gamma-ray emission measured during these IXPE observations.\par

The highest X-ray polarization degree of the campaign is $\approx 26\%$ on MJD~60290 (December 12 2023). It is close to the record value measured so far in an HSP, being $\approx 30\%$ in PKS~2155-304 \citep{2024arXiv240601693K}. The following day (MJD~60291, December 13 2023), the polarization was similarly high, $\sim 23\%$, and about 8 times higher than that in the optical. Simultaneous to this high polarization state, the multi-hour exposure from \textit{XMM-Newton} reveals an hysteresis spectral loop in counter-clockwise direction. Despite being a pattern observed in earlier \textit{XMM-Newton} observations of Mrk~421 and other HSPs \citep{2002ApJ...572..762Z, 2003A&A...402..929B, 2004A&A...424..841R}, it is the first time we can study it in combination with the recorded high X-ray polarization degree coupled to the strong chromatic behavior.\par 

Counter-clockwise loops are indicative of a delay from the high-energy X-ray band (here 2-10\,keV) with respect to the lower energy band (here 0.3-2\,keV). It can be explained by the gradual acceleration of particles towards higher energy, first reaching the 0.3-2\,keV radiation band and then the 2-10\,keV band. As discussed in \citet{1998A&A...333..452K}, such a high-energy delay becomes apparent when the radiation cooling timescale is comparable to the particle acceleration timescale, $t'_{\rm cool} \sim t'_{\rm acc}$, which occurs at the highest particle energy reachable by the system. The IXPE energy range, which largely overlaps with the 2-10\,keV band, therefore probes during that day the emission from particles that are located at (or very close to) the high-energy cutoff. In a shock acceleration scenario, the most energetic particles are located close to the shock front, and probe the highest degree of magnetic field ordering due to the compression of the helical field \citep{1985ApJ...298..114M, 2018MNRAS.480.2872T}. A high X-ray polarization degree is thus expected simultaneous to counter-clockwise hysteresis loops, which is supported by the data from our campaign.\par 

The polarization degree in the R-band simultaneous to the \textit{XMM-Newton} observation on MJD~60291 (December 13 2023) is drastically lower, by a factor $\approx 7$, and is in fact the largest difference observed in this campaign (see Fig.~\ref{fig:pdeg_ratios}). The lower polarization at decreasing frequencies is possibly caused by the advection and cooling of the freshly accelerated particles towards downstream regions of the jet where a higher degree of turbulence exists \citep{1985ApJ...298..114M, 2018MNRAS.480.2872T}. The TEMZ model developed by \citet{2014ApJ...780...87M}, which considers a number turbulent plasma cells crossing a standing conical shock, also predicts an increase of the polarization with frequency. It is caused by the higher acceleration efficiency in plasma cells whose magnetic field is almost parallel to the shock normal. Since the cells fulfilling such conditions are less numerous than those whose magnetic field is oriented away from the shock normal, the polarization increases at higher frequencies ($P_{\rm deg} \propto 1/ \sqrt{N}$, where $N$ is the number of cells). In such a situation, one also expects a stronger variability of the polarization at higher frequencies, in agreement with the observations reported here. Based on TEMZ simulations presented in \citet{2021Galax...9...27M} the X-ray polarization is expected to be on average $\approx 5\%$ higher (in absolute terms) than the one in the R-band. Although this chromaticity is lower than observed in December 2023 for several of the days, and in particular during the \textit{XMM-Newton} observations on MJD~60291 (December 13 2023). Nonetheless, since the counter-clockwise loops indicate that IXPE probed the emission from particles at the high-energy cutoff, we possibly probed an extreme case of the TEMZ model for which the chromatic trend may be more pronounced. This extreme situation would need to be investigated with a dedicated simulation to verify if TEMZ model is still in agreement with the data.\par 

\subsection{ Implications of the observed variability in the X-ray and VHE bands}

We find a $>3\sigma$ indication of a positive X-ray/VHE correlated variability (see Fig.~\ref{fig:xrt_vs_magic}). This is a general trend regularly reported for Mrk~421 and other HSPs \citep[see e.g.][]{2023ApJS..266...37A, 2021A&A...647A..88A, 2020A&A...638A..14M, 2009A&A...502..749A} that indicates a co-spatial origin and a common underlying population of radiating particles, supporting a leptonic origin of the emission. Nonetheless, a large scatter is measured in the correlation throughout the IXPE observing window. 
For a roughly equivalent X-ray flux, the VHE flux varies by almost a factor 2, implying an evolution with time of the Compton dominance (the ratio between the inverse-Compton and synchrotron peak luminosities). As discussed in \citet{2005A&A...433..479K}, the VHE/X-ray correlation in HSPs may exhibit very different trends (ranging from a sub-linear to more-than-quadratic relationship) depending on the parameters driving the variability. The observed scatter may thus be caused by simultaneous and uncorrelated changes in multiple characteristics, such as the size of the emitting region, the magnetic field, and the particle density varying simultaneously in an uncorrelated manner. \par   

We highlight that changes in the Compton dominance and a significant scatter in the VHE/X-ray correlation are naturally expected \citep[and have been observed, see e.g.][]{2021A&A...655A..89M} when considering datasets covering years or months. The reason is that, on such timescales, separate regions from the jet with different environment may dominate one after the other the broadband emission \citep[see e.g.][]{2019ApJ...877...26H}. In the present case, the evolution occurs down to daily timescales, which is much shorter. It further supports the scenario in which the flare is caused by a highly turbulent plasma crossing a shock front since its turbulent nature may easily explain simultaneous rapid (and stochastic) variations of multiple source environment parameters, that in turn produce a scatter in the correlation patterns.\par

Regarding the VHE and X-ray spectral characteristics, rather soft spectra are observed (see Sect.~\ref{sec:spectral_evolution}). For a given X-ray and VHE flux, the hardness ratio is on average lower compared to archival data. Mrk~421 commonly shows a harder-when-brighter behavior in VHE and X-rays \citep{2021MNRAS.504.1427A,2021A&A...655A..89M}, but this trend is only partly observed in the current campaign (see Fig.~\ref{fig:HR_xray_vhe}). While the hardness ratio over the entire period indeed shows a harder-when-brighter evolution, it becomes non-significant during the IXPE window. Focusing on the hardness ratio as function of the 0.3-2\,keV flux, the observations during the IXPE window show a clearly diverging pattern with respect to the rest of the campaign, and the spectral shape stays roughly constant. The absence of harder-when-brighter trend could again be indicative of a highly turbulent plasma crossing a shock front. For instance, if there were stochastic and uncorrelated changes in the electron density and the size of the emitting region, any harder-when-brighter trend would likely be washed out.\par

\subsection{Interpretation of the SED modelling within the context of a shock acceleration scenario}

We have modelled the evolution of the broadband SED across the IXPE window. The model consists of two overlapping components, one dominating the X-ray and VHE (the ``compact'' zone) and another dominating the UV/optical and MeV/GeV regimes (the ``extended'' zone). The radio band is dominated by the ``extended'' zone, but the model remains below the observed fluxes. We implicitly assumed that a significant fraction of the radio flux receives additional contribution from broader regions further downstream the jet. Both zones were assumed to be composed of $N$ turbulent plasma cells that have a roughly identical magnetic field strength but a random orientation, leading to an averaged observed polarization degree of $P_{\rm deg} \approx 75\% / \sqrt{N}$. The relative size of the components were then determined based on the optical-to-X-ray polarization degree. The ``compact'' zone is located close to the acceleration site and contains freshly accelerated particles, where the magnetic field is more ordered. When particles cool and advect, they populate a larger region (mimicked by the ``extended'' zone) where a higher degree of turbulences induces a drop in the polarization.\par 

The large X-ray polarization variability that is not observed in the optical, as well as the chromatic trend of the polarization degree,  indicate that the ``compact'' zone is largely subdominant in the optical regime. Using the magnetic field that we derived from the model ($\approx$ 0.05-0.08\,G), this implies a high minimum Lorentz factor $\gamma'_{\rm min} \gtrsim10^4$. This is significantly higher that the values used in previous modelling of BL Lac type object SEDs, which are usually $\gamma'_{\rm min} \sim 10^2$-$10^3$ \citep{2010MNRAS.401.1570T, 2016ApJ...819..156B}. In the case of extreme TeV BL Lac objects (that have an inverse-Compton peak located above 1\,TeV) values around $10^4$ are sometimes required to model the hard VHE spectrum \citep{2006MNRAS.368L..52K, 2018MNRAS.477.4257C}. In our present case, the high minimum Lorentz factor is robustly implied by the polarization data. A large minimum Lorentz factor is in fact expected in an electron-ion plasma. While in such a scenario most of the energy is carried by the ions, a fraction of it can be transferred to the electrons, leading to a large $\gamma'_{\rm min}$ in the electron population. In the case of relativistic shocks, particle-in-cell (PIC) simulations indicate that $\gamma'_{\rm electrons, min} \sim 600 \, \gamma_{\rm sh}$, where $\gamma_{\rm sh}$ is the shock velocity in the un-shocked plasma frame \citep{2011ApJ...726...75S, 2013ApJ...771...54S, 2021A&A...654A..96Z}. Hence, the data suggest accelerations at highly relativistic shocks ($\gamma_{\rm sh} \gtrsim 10$). The requirement of a relativistic shock is also in line with the relatively high Doppler factor ($\delta \sim 60$) needed to capture the X-ray/VHE spectrum.\par 

In the model, the cutoff Lorentz factor in the ``compact'' zone is fixed to $\gamma'_{\rm cutoff}= 1.3 \times 10^5$ for all days, and provides a good description of the X-ray/VHE SEDs. $\gamma'_{\rm cutoff}$ can not be precisely constrained with the data at hand, but it is constrained from below by the X-ray spectrum since values significantly lower than $1.3 \times 10^5$ can not accommodate the highest X-ray energy points. At $\gamma'_{\rm cutoff}= 1.3 \times 10^5$, the electrons radiate synchrotron photons in the $\approx$ 1-3\,keV band (according to the magnetic field value used in the model). Thus, in agreement with the hysteresis pattern in counter-clockwise direction discussed above, the X-ray data from IXPE, \textit{XMM-Newton} and \textit{Swift}-XRT are probing the cutoff energy of the accelerated electrons.\par

The temporal evolution of the SEDs can be well described by varying in both zones the electron density $N'_e$ and magnetic field $B'$, as well as as the slope of the injected electrons in the ``compact'' zone in order to capture the X-ray/VHE spectral variability. The electron density is the parameter that varies the most, up to a factor 3 (see Fig.~\ref{fig:parameter_evol}), while the magnetic fields changes by less than 40\%. We do not find any correlation between the fluxes and the evolution of any of the parameters, which again corroborates the stochastic nature of the physical mechanism driving the flare.\par

Our model shows that $B'$ within the ``compact'' zone tends to increase for smaller radii (see Fig.~\ref{fig:parameter_evol}), i.e. when the emitting particles are on average closer to the acceleration site. During the \textit{XMM-Newton} observation (MJD~60291, December 13 2023), the elevated X-ray polarization degree observed by IXPE leads to the smallest region radius among all the days based on the constrains described earlier ($R'=0.4 \times 10^{15}$\,cm), and in order to properly describe the data, the magnetic field reaches a maximum ($B'=0.08$\,G). We also find that the magnetic field is always stronger in the ``compact'' zone compared to that of the ``extended'' zone. PIC simulations show that the magnetic field at a shock front is expected to be self-amplified before decaying further downstream. The observed anti-correlation between $B'$ and $R'$ in our modelling is thus consistent with a shock-acceleration scenario \citep{2011ApJ...726...75S, 2018MNRAS.480.2872T}. Finally, we highlight that an anti-correlation between the radius and magnetic field is also expected to be caused by the cooling of the particles. Assuming that synchrotron radiation is the dominant cooling mechanisms, an increase of $B'$ will make the high-energy particles radiate their energy closer to the acceleration site, hence the size of the emitting region smaller. \par  

The slope of the electrons in the ``compact'' zone vary moderately throughout the days, $\max{(p)}-\min{(p)} = 0.45$. This is inline with the relatively moderate X-ray/VHE spectral variability discussed previously. We find that $p \approx$ 2.2-2.5 provides a good description of the majority of the X-ray and VHE spectra. Such slopes are close to the predictions from relativistic parallel shocks (i.e. $\approx 2.23$), and values close to $\approx 2.5$ can easily be produced, for instance, in case the shock is oblique instead of purely parallel \citep{2000ApJ...542..235K, 2012ApJ...745...63S}.\par

To conclude, this manuscript reports the first measurement of X-ray polarization from a HSP blazar during an X-ray and gamma-ray flaring activity, which has provided unprecedented information on the acceleration and radiation mechanisms driving blazar flares. The fast variations of the X-ray polarization, combined with an average angle remaining closely aligned with the one of the jet (and in the optical), favours a scenario in which the flare is caused by a highly turbulent electron-ion plasma crossing the shock front where the particles get accelerated. The measured increase of the polarization degree with frequency further implies that the emitting region is stratified, and that the most energetic particles are located closest to the shock. The multi-wavelength spectral properties, intra-band correlation, and the counter-clockwise loop in the hardness ratio of the X-ray emission supports such scenario, which can also be described, in first order,  within a theoretical leptonic model based on two components. Further multi-wavelength observations that include IXPE during flaring episodes of blazars will be crucial to determine if the observed trends are common characteristics of blazar flares.

\section*{Author contribution}

A. Arbet Engels: project management, P.I. of MAGIC observations, organization of multi-wavelength observations and data analysis, MAGIC analysis, correlation analysis, theoretical modeling and interpretation, paper drafting; I. Liodakis: organization of multi-wavelength observations and data analysis; L. Heckmann: project management, MAGIC and \textit{Fermi}-LAT data analysis, variability analysis, theoretical interpretation, paper drafting; D. Paneque: organization of the MWL observations, theoretical interpretation, paper drafting; The rest of the authors have contributed in one or several of the following ways: design, construction, maintenance and operation of the instrument(s) used to acquire the data; preparation and/or evaluation of the observation proposals; data acquisition, processing, calibration and/or reduction; production of analysis tools and/or related Monte Carlo simulations; overall discussions about the contents of the draft, as well as related refinements in the descriptions.

\begin{acknowledgements}
We would like to thank the Instituto de Astrof\'{\i}sica de Canarias for the excellent working conditions at the Observatorio del Roque de los Muchachos in La Palma. The financial support of the German BMBF, MPG and HGF; the Italian INFN and INAF; the Swiss National Fund SNF; the grants PID2019-104114RB-C31, PID2019-104114RB-C32, PID2019-104114RB-C33, PID2019-105510GB-C31, PID2019-107847RB-C41, PID2019-107847RB-C42, PID2019-107847RB-C44, PID2019-107988GB-C22, PID2022-136828NB-C41, PID2022-137810NB-C22, PID2022-138172NB-C41, PID2022-138172NB-C42, PID2022-138172NB-C43, PID2022-139117NB-C41, PID2022-139117NB-C42, PID2022-139117NB-C43, PID2022-139117NB-C44 funded by the Spanish MCIN/AEI/ 10.13039/501100011033 and “ERDF A way of making Europe”; the Indian Department of Atomic Energy; the Japanese ICRR, the University of Tokyo, JSPS, and MEXT; the Bulgarian Ministry of Education and Science, National RI Roadmap Project DO1-400/18.12.2020 and the Academy of Finland grant nr. 320045 is gratefully acknowledged. This work was also been supported by Centros de Excelencia ``Severo Ochoa'' y Unidades ``Mar\'{\i}a de Maeztu'' program of the Spanish MCIN/AEI/ 10.13039/501100011033 (CEX2019-000920-S, CEX2019-000918-M, CEX2021-001131-S) and by the CERCA institution and grants 2021SGR00426 and 2021SGR00773 of the Generalitat de Catalunya; by the Croatian Science Foundation (HrZZ) Project IP-2022-10-4595 and the University of Rijeka Project uniri-prirod-18-48; by the Deutsche Forschungsgemeinschaft (SFB1491) and by the Lamarr-Institute for Machine Learning and Artificial Intelligence; by the Polish Ministry Of Education and Science grant No. 2021/WK/08; and by the Brazilian MCTIC, CNPq and FAPERJ.

The \textit{Fermi} LAT Collaboration acknowledges generous ongoing support from a number of agencies and institutes that have supported both the development and the operation of the LAT as well as scientific data analysis. These include the National Aeronautics and Space Administration and the Department of Energy in the United States, the Commissariat \`a l'Energie Atomique and the Centre National de la Recherche Scientifique / Institut National de Physique Nucl\'eaire et de Physique des Particules in France, the Agenzia Spaziale Italiana and the Istituto Nazionale di Fisica Nucleare in Italy, the Ministry of Education, Culture, Sports, Science and Technology (MEXT), High Energy Accelerator Research Organization (KEK) and Japan Aerospace Exploration Agency (JAXA) in Japan, and the K.~A.~Wallenberg Foundation, the Swedish Research Council and the Swedish National Space Board in Sweden. Additional support for science analysis during the operations phase is gratefully acknowledged from the Istituto Nazionale di Astrofisica in Italy and the Centre National d'\'Etudes Spatiales in France. This work performed in part under DOE Contract DE-AC02-76SF00515.

The corresponding authors of this manuscript, namely Axel Arbet-Engels, Lea Heckmann and David Paneque, acknowledge support from the Deutsche Forschungs gemeinschaft (DFG, German Research Foundation) under Germany’s Excellence Strategy – EXC-2094 – 390783311.

The Imaging X-ray Polarimetry Explorer ({\it IXPE}) is a joint US and Italian mission. 
The US contribution is supported by the National Aeronautics and Space Administration (NASA) and led and managed by its Marshall Space Flight Center (MSFC), with industry partner Ball Aerospace (contract NNM15AA18C). The Italian contribution is supported by the Italian Space Agency (Agenzia Spaziale Italiana, ASI) through contract ASI-OHBI-2017-12-I.0, agreements ASI-INAF-2017-12-H0 and ASI-INFN-2017.13-H0, and its Space Science Data Center (SSDC), and by the Istituto Nazionale di Astrofisica (INAF) and the Istituto Nazionale di Fisica Nucleare (INFN) in Italy. This research used data products provided by the {\it IXPE} Team (MSFC, SSDC, INAF, and INFN) and distributed with additional software tools by the High-Energy Astrophysics Science Archive Research Center (HEASARC), at NASA Goddard Space Flight Center (GSFC).

I. L. and S.K. were funded by the European Union ERC-2022-STG - BOOTES - 101076343. Views and opinions expressed are however those of the author(s) only and do not necessarily reflect those of the European Union or the European Research Council Executive Agency. Neither the European Union nor the granting authority can be held responsible for them

Some of the data are based on observations collected at the Observatorio de Sierra Nevada; which is owned and operated by the Instituto de Astrof\'isica de Andaluc\'ia (IAA-CSIC); and at the Centro Astron\'{o}mico Hispano en Andalucía (CAHA); which is operated jointly by Junta de Andaluc\'{i}a and Consejo Superior de Investigaciones Cient\'{i}ficas (IAA-CSIC). 

The research at Boston University was supported in part by National Science Foundation grant AST-2108622, NASA Fermi Guest Investigator grants 80NSSC21K1917 and 80NSSC22K1571, and NASA Swift Guest Investigator grant 80NSSC22K0537. This research was conducted in part using the Mimir instrument, jointly developed at Boston University and Lowell Observatory and supported by NASA, NSF, and the W.M. Keck Foundation. We thank D.~Clemens for guidance in the analysis of the Mimir data. 

This study used observations conducted with the 1.8m Perkins Telescope (PTO) in Arizona (USA), which is owned and operated by Boston University. This work was supported by NSF grant AST-2109127. 

We acknowledge the use of public data from the Swift data archive. Based on observations obtained with XMM-Newton, an ESA science mission with instruments and contributions directly funded by ESA Member States and NASA.

This work has made use of data from the Joan Oró Telescope (TJO) of the Montsec Observatory (OdM), which is owned by the Catalan Government and operated by the Institute for Space Studies of Catalonia (IEEC).

The Submillimeter Array is a joint project between the Smithsonian Astrophysical Observatory and the Academia Sinica Institute of Astronomy and Astrophysics and is funded by the Smithsonian Institution and the Academia Sinica. We recognize that Maunakea is a culturally important site for the indigenous Hawaiian people; we are privileged to study the cosmos from its summit.

The 100 m radio telescope at Effelsberg is operated by the Max-Planck-Institut für Radioastronomie (MPIfR) on behalf of the Max-Planck-Society. Observations with the 100 m telescope have received funding from the European Union's Horizon 2020 research and innovation program under grant agreement No. 101004719 (ORP).

The Liverpool Telescope is operated on the island of La Palma by Liverpool John Moores University in the Spanish Observatorio del Roque de los Muchachos of the Instituto de Astrofisica de Canarias with financial support from the UK Science and Technology Facilities Council.

\end{acknowledgements}

\bibliographystyle{aa}
\bibliography{bibliography_paper.bib}

\begin{appendix}


\section{\textit{Swift}-XRT spectral fits}
\label{sec:spectral_fits_xrt}
The spectral fits obtained for each \textit{Swift}-XRT observation are presented in Table~\ref{tab:swift_spectral_param}.
\begin{table*}
\caption{\label{tab:swift_spectral_param}\textit{Swift}-XRT spectral analysis results.} 
\centering
\begin{tabular}{ l c c c c c c c c c c}     
\hline\hline 
 MJD & $F_{0.3-2\text{\,keV}}$ & $F_{2-10\text{\,keV}}$ & $\Gamma$ & $\chi^2$/dof & $\alpha$ & $\beta$ & $\chi^2$/dof \\
  & $[10^{-11} \mathrm{erg} \, \mathrm{cm}^{-2} \mathrm{s}^{-1}]$ & $[10^{-11} \mathrm{erg}\, \mathrm{cm}^{-2} \mathrm{s}^{-1}]$ &  &  &  &  &\\
\hline\hline  
60259.23 & 86.91 $\pm$ 0.66 & 48.58 $\pm$ 0.89 & 2.21 $\pm$ 0.01 & 426/358 & 2.15 $\pm$ 0.01 & 0.18 $\pm$ 0.03 & 384/357\\
60261.21 & 73.89 $\pm$ 0.56 & 40.72 $\pm$ 0.86 & 2.20 $\pm$ 0.01 & 412/320 & 2.15 $\pm$ 0.01 & 0.21 $\pm$ 0.03 & 361/319\\
60263.18 & 80.56 $\pm$ 0.66 & 40.77 $\pm$ 0.77 & 2.25 $\pm$ 0.01 & 580/347 & 2.14 $\pm$ 0.01 & 0.33 $\pm$ 0.03 & 446/346\\
60265.23 & 57.30 $\pm$ 0.73 & 17.18 $\pm$ 0.57 & 2.54 $\pm$ 0.02 & 239/215 & 2.50 $\pm$ 0.02 & 0.23 $\pm$ 0.05 & 218/214\\
60267.21 & 84.47 $\pm$ 0.60 & 39.12 $\pm$ 0.72 & 2.30 $\pm$ 0.01 & 462/336 & 2.23 $\pm$ 0.01 & 0.25 $\pm$ 0.03 & 383/335\\
60269.26 & 59.39 $\pm$ 0.60 & 21.25 $\pm$ 0.58 & 2.44 $\pm$ 0.01 & 362/262 & 2.37 $\pm$ 0.02 & 0.28 $\pm$ 0.04 & 306/261\\
60271.24 & 55.67 $\pm$ 0.69 & 17.72 $\pm$ 0.61 & 2.49 $\pm$ 0.02 & 264/211 & 2.42 $\pm$ 0.02 & 0.33 $\pm$ 0.04 & 223/210\\
60284.76 & 74.45 $\pm$ 0.69 & 21.91 $\pm$ 0.55 & 2.53 $\pm$ 0.01 & 325/262 & 2.46 $\pm$ 0.01 & 0.35 $\pm$ 0.04 & 233/261\\
60286.01 & 77.73 $\pm$ 0.79 & 25.14 $\pm$ 0.68 & 2.48 $\pm$ 0.01 & 336/247 & 2.40 $\pm$ 0.02 & 0.37 $\pm$ 0.05 & 260/246\\
60288.19 & 70.53 $\pm$ 0.66 & 23.35 $\pm$ 0.56 & 2.48 $\pm$ 0.01 & 324/270 & 2.43 $\pm$ 0.01 & 0.25 $\pm$ 0.04 & 275/269\\
60290.51 & 83.48 $\pm$ 1.01 & 44.41 $\pm$ 1.34 & 2.24 $\pm$ 0.01 & 265/243 & 2.21 $\pm$ 0.02 & 0.12 $\pm$ 0.05 & 258/242\\
60291.09 & 117.11 $\pm$ 1.16 & 44.97 $\pm$ 0.97 & 2.40 $\pm$ 0.01 & 365/289 & 2.35 $\pm$ 0.01 & 0.24 $\pm$ 0.03 & 312/288\\
60291.23 & 101.03 $\pm$ 0.65 & 42.46 $\pm$ 0.72 & 2.35 $\pm$ 0.01 & 571/359 & 2.26 $\pm$ 0.01 & 0.32 $\pm$ 0.03 & 418/358\\
60291.50 & 71.01 $\pm$ 0.65 & 28.85 $\pm$ 0.64 & 2.37 $\pm$ 0.01 & 366/282 & 2.29 $\pm$ 0.01 & 0.29 $\pm$ 0.03 & 290/281\\
60292.08 & 93.50 $\pm$ 0.73 & 38.19 $\pm$ 0.72 & 2.37 $\pm$ 0.01 & 414/327 & 2.30 $\pm$ 0.01 & 0.27 $\pm$ 0.03 & 320/326\\
60292.21 & 109.12 $\pm$ 1.00 & 47.57 $\pm$ 1.02 & 2.33 $\pm$ 0.01 & 392/296 & 2.24 $\pm$ 0.01 & 0.31 $\pm$ 0.03 & 306/295\\
60292.47 & 116.69 $\pm$ 0.77 & 49.80 $\pm$ 0.83 & 2.34 $\pm$ 0.01 & 632/360 & 2.25 $\pm$ 0.01 & 0.31 $\pm$ 0.03 & 470/359\\
60294.19 & 95.92 $\pm$ 0.82 & 36.27 $\pm$ 0.69 & 2.41 $\pm$ 0.01 & 491/315 & 2.32 $\pm$ 0.01 & 0.32 $\pm$ 0.03 & 362/314\\
60294.52 & 106.21 $\pm$ 0.77 & 40.76 $\pm$ 0.70 & 2.40 $\pm$ 0.01 & 480/342 & 2.33 $\pm$ 0.01 & 0.29 $\pm$ 0.03 & 351/341\\
60295.18 & 128.41 $\pm$ 0.88 & 50.23 $\pm$ 0.83 & 2.39 $\pm$ 0.01 & 603/356 & 2.31 $\pm$ 0.01 & 0.29 $\pm$ 0.03 & 468/355\\
60295.25 & 127.24 $\pm$ 0.88 & 48.99 $\pm$ 0.89 & 2.39 $\pm$ 0.01 & 559/351 & 2.31 $\pm$ 0.01 & 0.33 $\pm$ 0.03 & 391/350\\
60296.17 & 125.86 $\pm$ 0.99 & 41.34 $\pm$ 0.74 & 2.48 $\pm$ 0.01 & 519/338 & 2.40 $\pm$ 0.01 & 0.33 $\pm$ 0.03 & 345/337\\
60296.51 & 111.52 $\pm$ 0.87 & 36.80 $\pm$ 0.71 & 2.49 $\pm$ 0.01 & 472/322 & 2.43 $\pm$ 0.01 & 0.26 $\pm$ 0.03 & 383/321\\
60298.16 & 93.39 $\pm$ 0.67 & 35.85 $\pm$ 0.67 & 2.40 $\pm$ 0.01 & 405/321 & 2.35 $\pm$ 0.01 & 0.24 $\pm$ 0.03 & 338/320\\
60298.61 & 107.15 $\pm$ 0.77 & 44.51 $\pm$ 0.77 & 2.36 $\pm$ 0.01 & 521/350 & 2.29 $\pm$ 0.01 & 0.26 $\pm$ 0.03 & 421/349\\
60299.20 & 125.62 $\pm$ 1.16 & 40.68 $\pm$ 0.73 & 2.51 $\pm$ 0.01 & 529/332 & 2.41 $\pm$ 0.01 & 0.33 $\pm$ 0.03 & 386/331\\
60301.24 & 134.57 $\pm$ 0.99 & 58.34 $\pm$ 0.92 & 2.33 $\pm$ 0.01 & 510/369 & 2.26 $\pm$ 0.01 & 0.28 $\pm$ 0.03 & 358/368\\
60315.11 & 61.90 $\pm$ 0.67 & 23.45 $\pm$ 0.67 & 2.41 $\pm$ 0.01 & 305/251 & 2.36 $\pm$ 0.02 & 0.22 $\pm$ 0.04 & 276/250\\
60317.09 & 36.36 $\pm$ 0.61 & 8.47 $\pm$ 0.36 & 2.69 $\pm$ 0.02 & 246/184 & 2.61 $\pm$ 0.02 & 0.33 $\pm$ 0.06 & 212/183\\
60319.06 & 47.78 $\pm$ 0.75 & 17.49 $\pm$ 0.49 & 2.44 $\pm$ 0.01 & 325/255 & 2.37 $\pm$ 0.02 & 0.25 $\pm$ 0.04 & 289/254\\
60321.18 & 65.70 $\pm$ 0.78 & 25.57 $\pm$ 0.69 & 2.39 $\pm$ 0.01 & 293/254 & 2.34 $\pm$ 0.02 & 0.23 $\pm$ 0.04 & 260/253\\
60323.16 & 92.19 $\pm$ 0.70 & 37.78 $\pm$ 0.73 & 2.36 $\pm$ 0.01 & 493/327 & 2.29 $\pm$ 0.01 & 0.29 $\pm$ 0.03 & 387/326\\

\hline 
\end{tabular}
\tablefoot{For each observation, the start time in MJD is given as well as the 0.3-2\,keV and 2-10\,keV fluxes. The observations have a typical exposure of about 1\,ks. The best-fit power-law indices $\Gamma$ are listed in the fifth column with the corresponding $\chi^2$/dof in the sixth column. The best-fit parameters $\alpha$ and $\beta$ from the log-parabolic fits with a pivot energy fixed at 1\,keV are also given with their corresponding $\chi^2$/dof. The analysis includes a photoelectric absorption by a fixed column density of $N_{\rm H}=1.34\times 10^{20}$\,cm$^{-2}$ \citep{2016A&A...594A.116H}.} 
\end{table*}

\section{MAGIC spectral fits}
\label{sec:spectral_fits_magic}
The spectral fits obtained for the nightly binned MAGIC observations are presented in Table~\ref{tab:MAGIC_spectral_param}.
\begin{table*}
\caption{\label{tab:MAGIC_spectral_param}MAGIC spectral analysis results.} 
\centering
\begin{tabular}{ l c c c c c c c }     
\hline\hline 
 MJD & $F_{0.2-1\text{\,TeV}}$ & $F_{>1\text{\,TeV}}$ & $f_0$ & $\alpha$ & $\chi^2$/dof\\  
 &   $[10^{-10} \mathrm{cm}^{-2} \mathrm{s}^{-1}]$ & $[10^{-11} \mathrm{cm}^{-2} \mathrm{s}^{-1}]$ & $[10^{-10} \mathrm{cm}^{-2} \mathrm{s}^{-1} \mathrm{TeV}^{-1}]$ &  &  \\
\hline\hline   
60259.2 & 2.73 $\pm$ 0.17 & 3.16 $\pm$ 0.30 & 7.81 $\pm$ 0.55 & 1.77 $\pm$ 0.07 & 5/11 \\
60261.2 & 2.93 $\pm$ 0.18 & 2.95 $\pm$ 0.30 & 8.20 $\pm$ 0.58 & 1.88 $\pm$ 0.07 & 21/11 \\
60263.2 & 2.30 $\pm$ 0.15 & 1.30 $\pm$ 0.23 & 7.38 $\pm$ 0.57 & 2.26 $\pm$ 0.11 & 6/9 \\
60267.2 & 2.52 $\pm$ 0.15 & 1.59 $\pm$ 0.22 & 7.85 $\pm$ 0.52 & 2.10 $\pm$ 0.08 & 9/10 \\
60271.2 & 1.51 $\pm$ 0.11 & 0.69 $\pm$ 0.14 & 5.16 $\pm$ 0.40 & 2.46 $\pm$ 0.12 & 5/7 \\
60287.2 & 3.05 $\pm$ 0.19 & 1.80 $\pm$ 0.23 & 9.51 $\pm$ 0.61 & 2.14 $\pm$ 0.07 & 16/11 \\
60288.2 & 1.52 $\pm$ 0.08 & 0.91 $\pm$ 0.10 & 5.04 $\pm$ 0.29 & 2.26 $\pm$ 0.08 & 19/9 \\
60290.2 & 2.74 $\pm$ 0.16 & 2.08 $\pm$ 0.27 & 8.71 $\pm$ 0.57 & 2.07 $\pm$ 0.08 & 16/8 \\
60291.2 & 3.66 $\pm$ 0.09 & 2.40 $\pm$ 0.11 & 11.55 $\pm$ 0.30 & 2.18 $\pm$ 0.03 & 6/10 \\
60292.2 & 3.52 $\pm$ 0.10 & 2.42 $\pm$ 0.12 & 10.87 $\pm$ 0.32 & 2.21 $\pm$ 0.04 & 14/11 \\
60294.2 & 2.89 $\pm$ 0.11 & 1.61 $\pm$ 0.13 & 9.65 $\pm$ 0.36 & 2.27 $\pm$ 0.05 & 14/11 \\
60295.2 & 3.86 $\pm$ 0.13 & 1.75 $\pm$ 0.15 & 12.33 $\pm$ 0.43 & 2.35 $\pm$ 0.05 & 20/9 \\
60296.2 & 3.01 $\pm$ 0.16 & 1.05 $\pm$ 0.19 & 9.89 $\pm$ 0.57 & 2.48 $\pm$ 0.09 & 7/8 \\
60299.2 & 4.62 $\pm$ 0.24 & 2.30 $\pm$ 0.33 & 15.64 $\pm$ 0.86 & 2.30 $\pm$ 0.08 & 5/7 \\
60315.1 & 1.88 $\pm$ 0.15 & 0.70 $\pm$ 0.18 & 6.34 $\pm$ 0.56 & 2.49 $\pm$ 0.14 & 4/8 \\
60319.2 & 1.07 $\pm$ 0.13 & 0.34 $\pm$ 0.15 & 3.52 $\pm$ 0.47 & 2.40 $\pm$ 0.21 & 5/6 \\

\hline 
\end{tabular}
\tablefoot{The observing day is listed in the first column. The 0.2-1\,keV and >1\,TeV fluxes are given in the second and third column. The amplitude $f_0$ and index $\alpha$ of the log-parabolic fit are given in the third and fourth column (the normalization energy is 300\,GeV). Their corresponding $\chi^2/dof$ are listed in the fifth column. All fits are performed after fixing the curvature parameter to $\beta=0.5$ (see Sect.~\ref{sec:spectral_evolution}) and correcting for the EBL absorption using the model from \citet{2011MNRAS.410.2556D}.} 
\end{table*}

\section{X-ray polarization correlated with the flux spectral hardness measured in the X-ray and VHE bands}
\label{sec:xray_pol_corr}

\begin{figure*}
    \centering
    \begin{subfigure}[b]{0.497\textwidth}
        \centering
        \includegraphics[width=\textwidth]{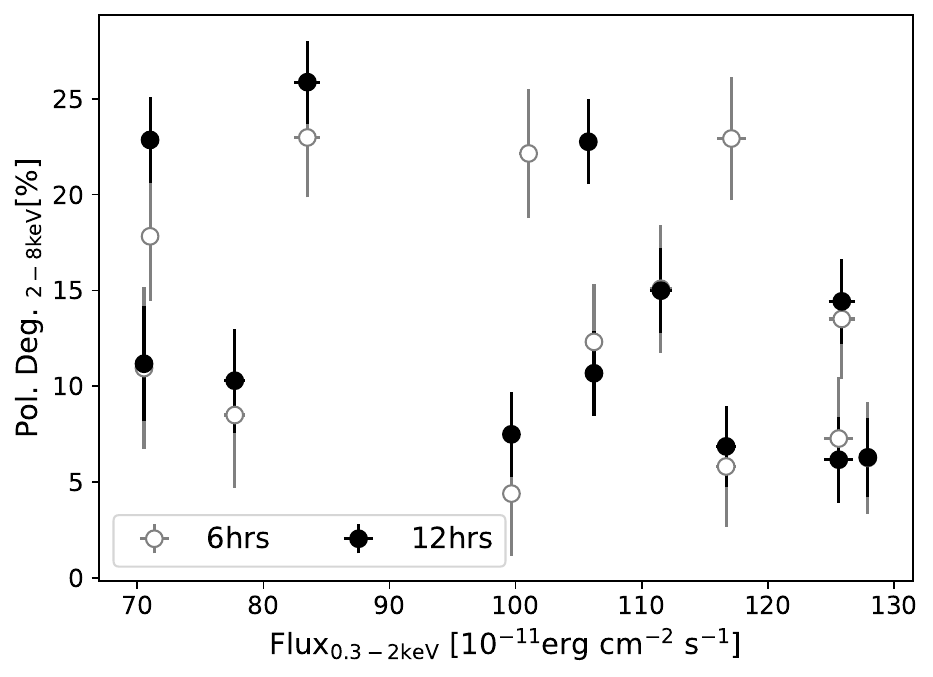}
    \end{subfigure}
    \begin{subfigure}[b]{0.497\textwidth}  
        \centering 
        \includegraphics[width=\textwidth]{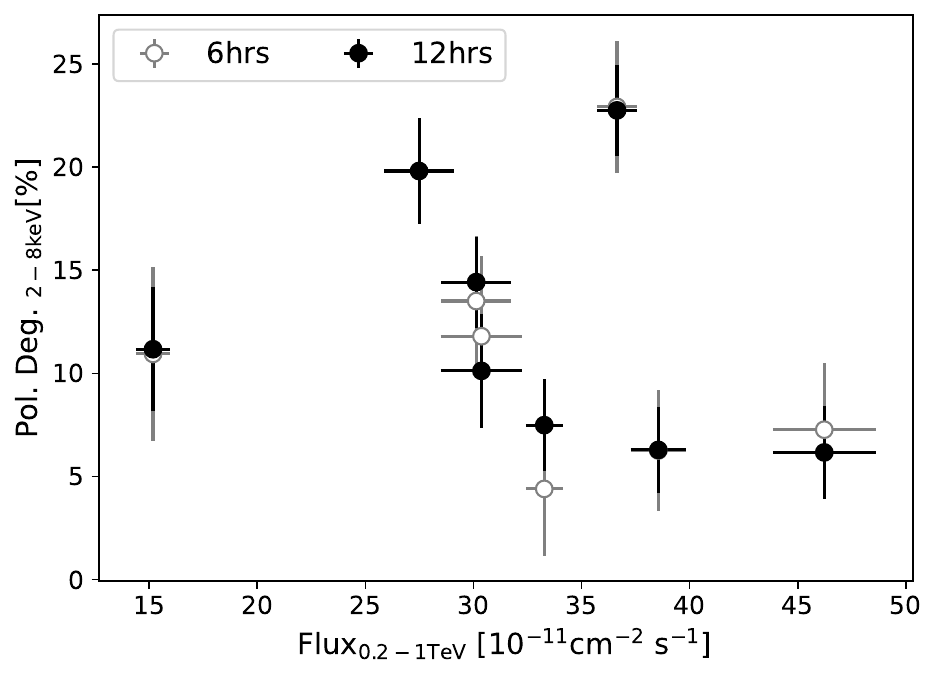}
    \end{subfigure}
    \begin{subfigure}[b]{0.497\textwidth}
        \centering
        \includegraphics[width=\textwidth]{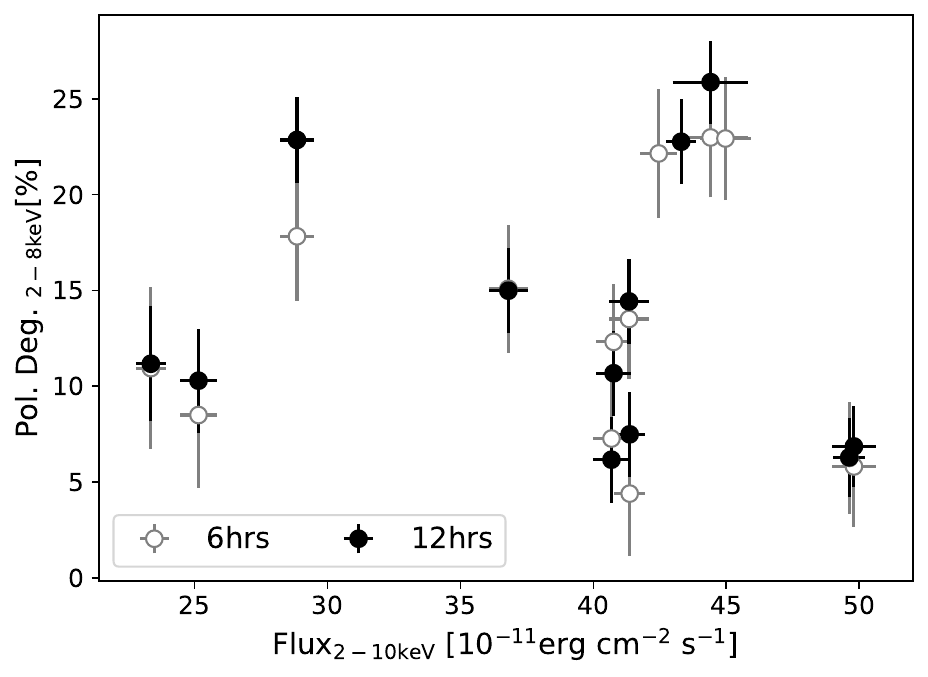}
    \end{subfigure}
    \begin{subfigure}[b]{0.497\textwidth}  
        \centering 
        \includegraphics[width=\textwidth]{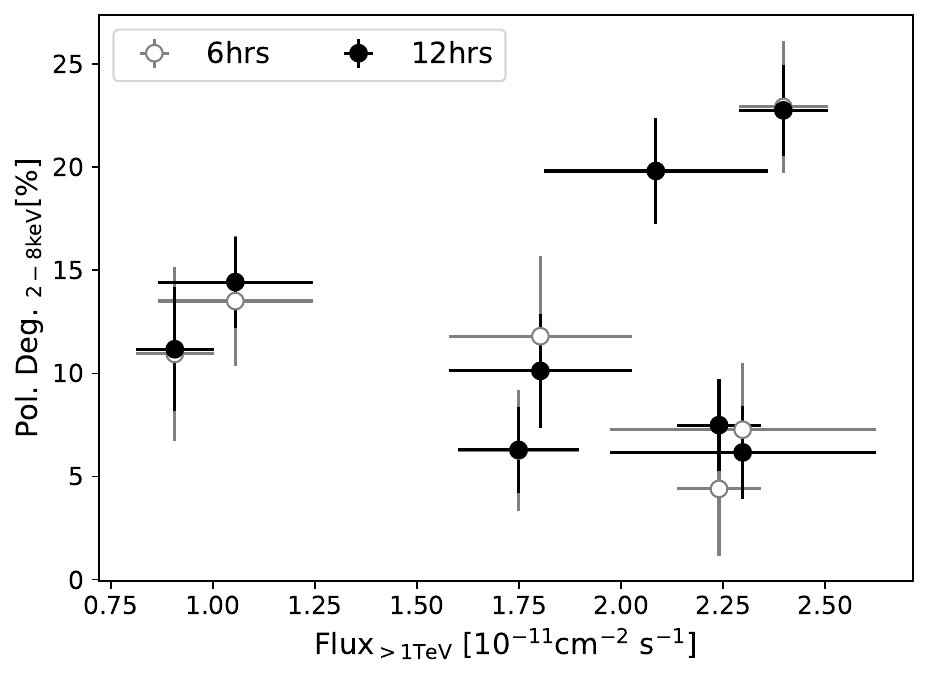}
    \end{subfigure}
    \begin{subfigure}[b]{0.497\textwidth}  
        \centering 
        \includegraphics[width=\textwidth]{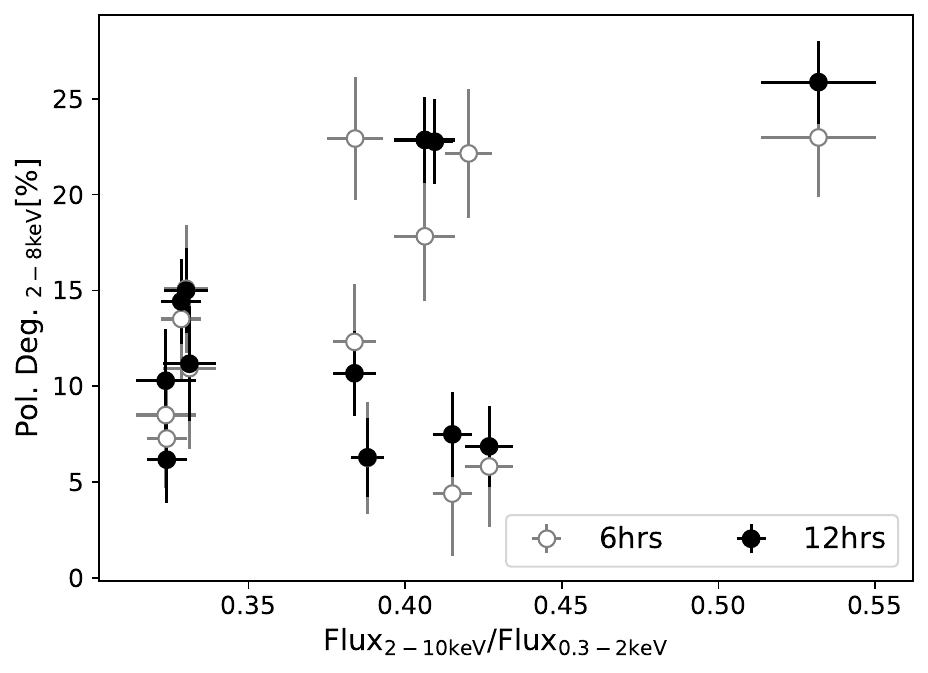}
    \end{subfigure}
    \begin{subfigure}[b]{0.497\textwidth}  
        \centering 
        \includegraphics[width=\textwidth]{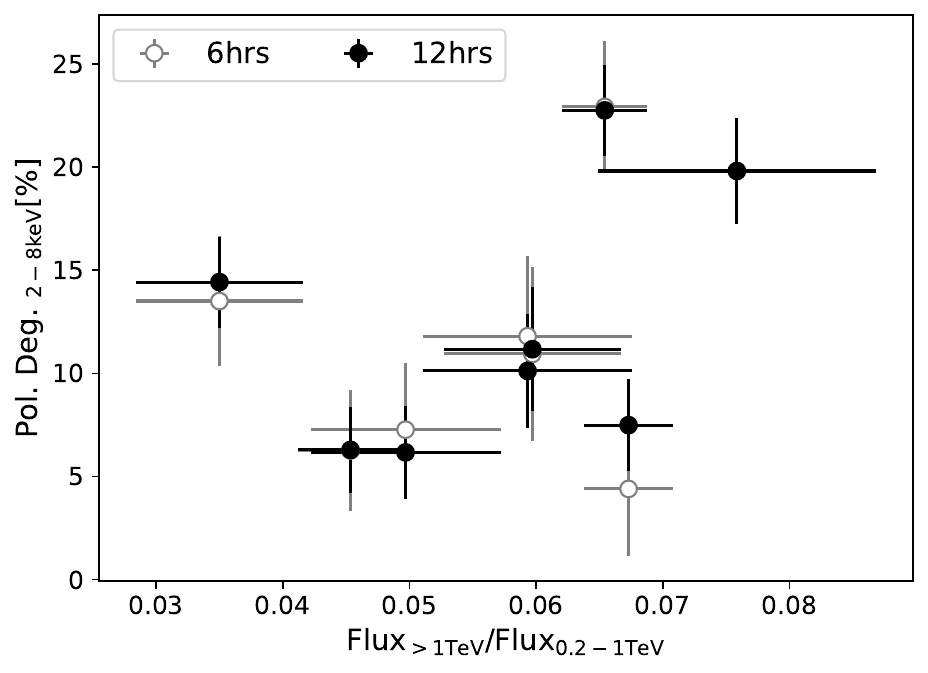}
    \end{subfigure}
    \caption{IXPE X-ray polarization degree correlated with the simultaneous flux and hardness ratio measurements in the different X-ray and VHE energy bands from \textit{Swift}-XRT and MAGIC. Black markers show results using IXPE data binned over 12\,hrs, empty grey markers IXPE data binned over 6\,hrs.}
    \label{fig:pdeg_vs_fluxes}
\end{figure*}

\begin{figure*}
    \centering
    \begin{subfigure}[b]{0.497\textwidth}
        \centering
        \includegraphics[width=\textwidth]{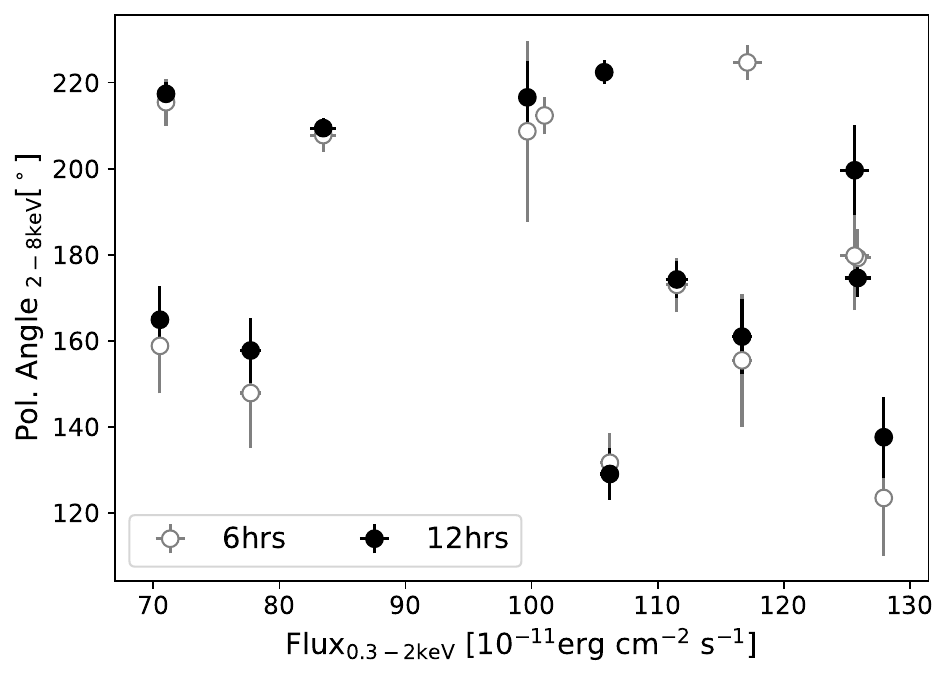}
    \end{subfigure}
    \begin{subfigure}[b]{0.497\textwidth}  
        \centering 
        \includegraphics[width=\textwidth]{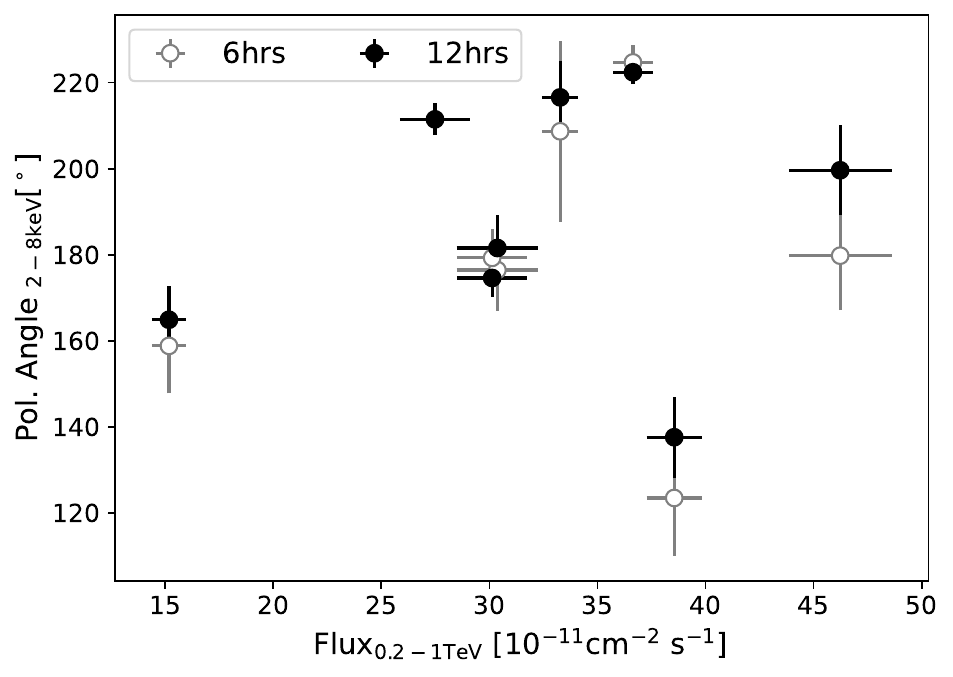}
    \end{subfigure}
    \begin{subfigure}[b]{0.497\textwidth}
        \centering
        \includegraphics[width=\textwidth]{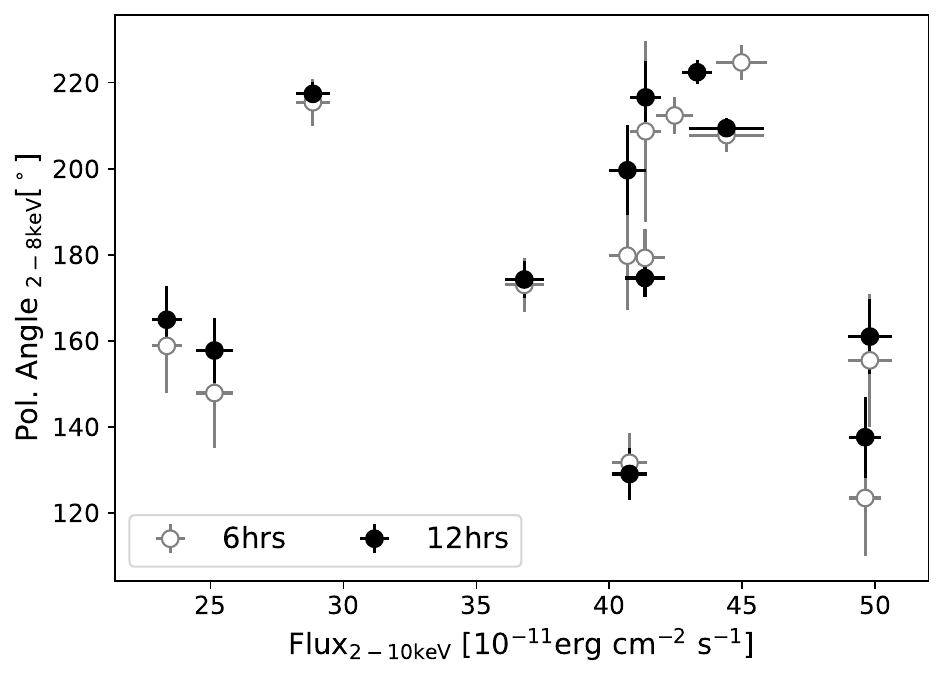}
    \end{subfigure}
    \begin{subfigure}[b]{0.497\textwidth}  
        \centering 
        \includegraphics[width=\textwidth]{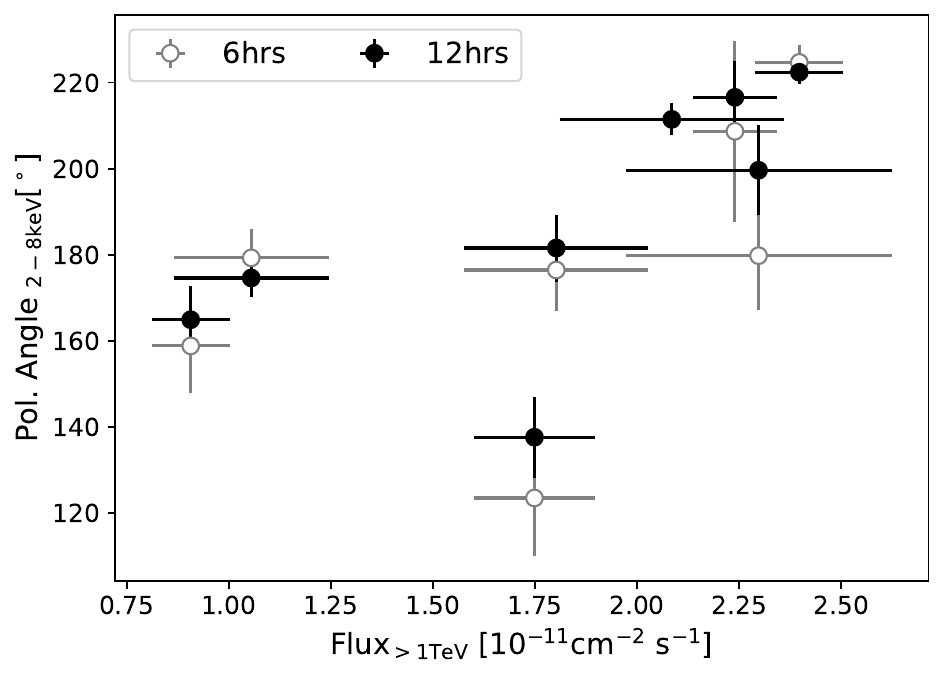}
    \end{subfigure}
    \begin{subfigure}[b]{0.497\textwidth}  
        \centering 
        \includegraphics[width=\textwidth]{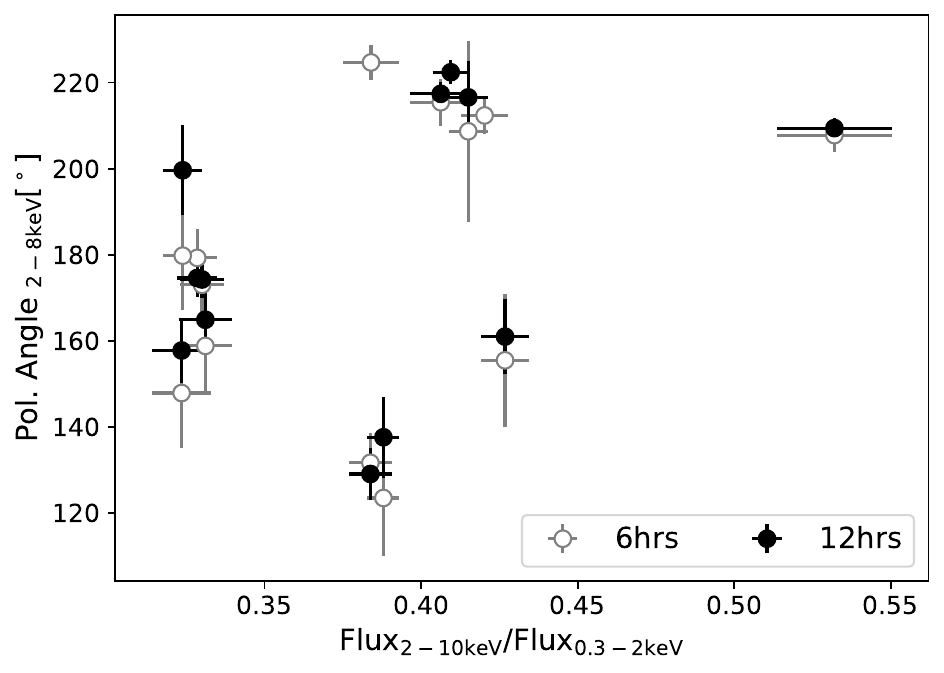}
    \end{subfigure}
    \begin{subfigure}[b]{0.497\textwidth}  
        \centering 
        \includegraphics[width=\textwidth]{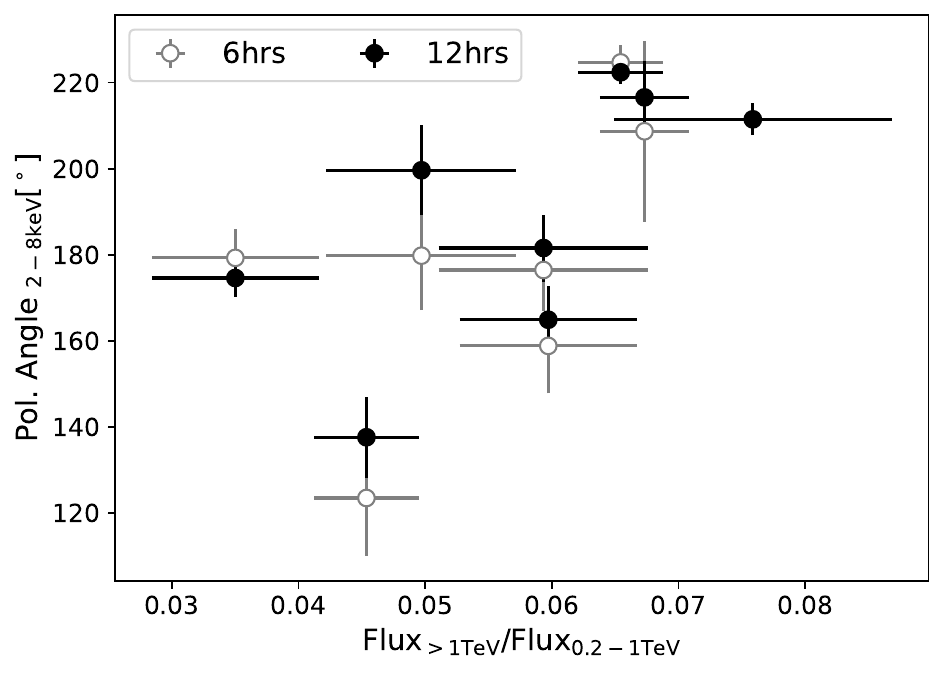}
    \end{subfigure}
    \caption{IXPE X-ray polarization angle correlated with the simultaneous flux and hardness ratio measurements in the different X-ray and VHE energy bands from \textit{Swift}-XRT and MAGIC. Black markers show results using IXPE data binned over 12\,hrs, empty grey markers IXPE data binned over 6\,hrs.}
    \label{fig:pa_vs_fluxes}
\end{figure*}

\begin{figure}
    \centering 
    \includegraphics[width=0.497\textwidth]{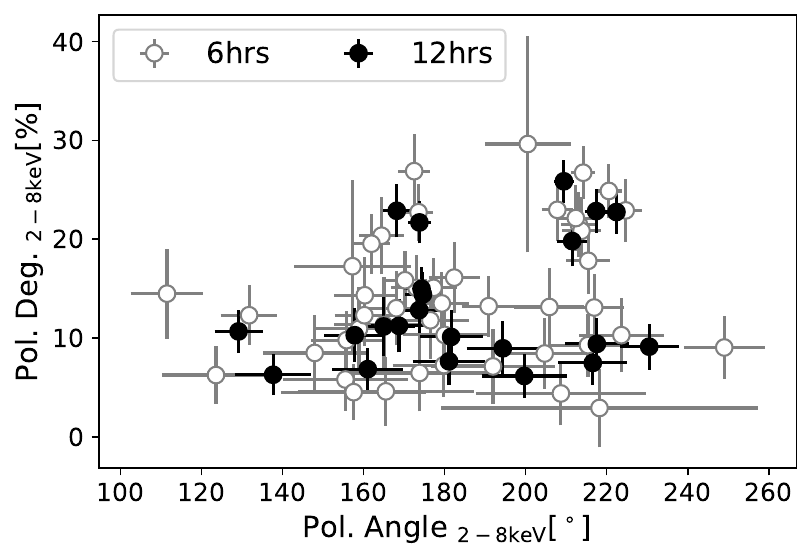}
    \caption{Comparison of the IXPE X-ray polarization degree and angle. Black markers show results using IXPE data binned over 12\,hrs, empty grey markers IXPE data binned over 6\,hrs.}
    \label{fig:pdeg_vs_pa_ixpe}
\end{figure}

This section summarizes the results on a search for possible correlation between the X-ray polarization and the flux observed in the X-ray and VHE bands. The correlation was conducted using the IXPE data binned into 6\,hrs and 12\,hrs intervals as presented in Fig.~\ref{fig:longterm_LC}. In Fig.~\ref{fig:pdeg_vs_fluxes}, the X-ray polarization degree is correlated with the low-energy (top-panels) and high-energy (mid panels) bands of the X-ray and VHE data. In the bottom panels the correlation is further performed using the hardness ratios in each energy regimes. In Fig.~\ref{fig:pa_vs_fluxes} we performed a similar analysis but using the X-ray polarization angle. Finally, we also correlated the X-ray polarization degree and angle between each other, and the results are shown in Fig.~\ref{fig:pdeg_vs_pa_ixpe}. In all the cases, no significant ($>3\sigma$) correlation is found.

\section{SED modelling with a lower Doppler factor}
\label{sec:modelling_30_test}
\begin{figure}[h!]
    \centering
    \includegraphics[width=1\columnwidth]{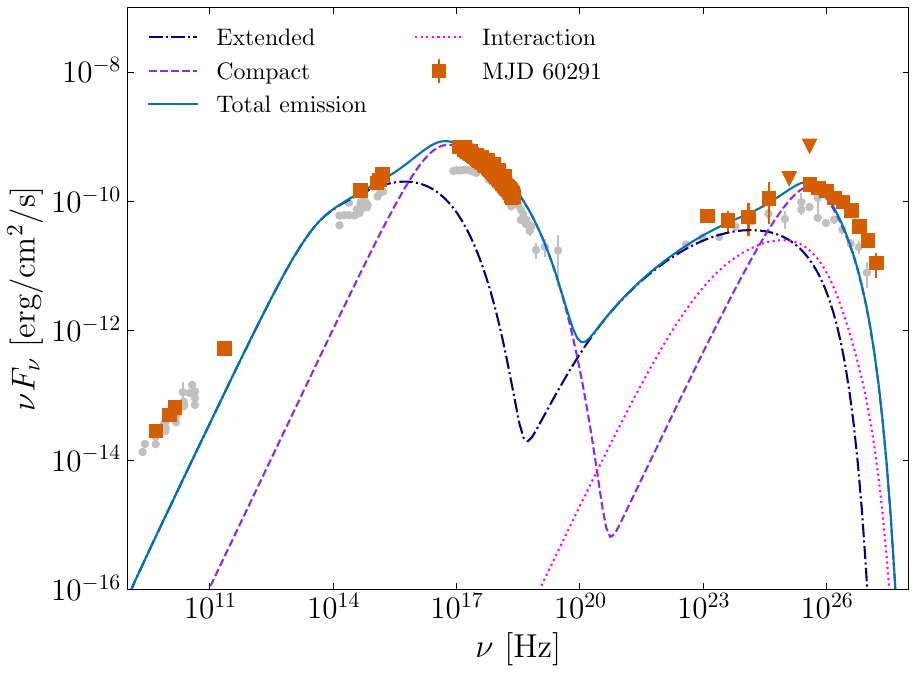}
    \caption{Modelling of the broadband SED from MJD~60291 using $\delta=30$.} 
    \label{fig:modelling_test_delta30}
\end{figure}

In the modelling presented in Sect.~\ref{sec:modelling}, we argued that a Doppler factor as large as $\delta=60$ is necessary to properly capture the measured X-ray and the VHE spectra. With a lower value around $\delta=30$, as typically used in archival works, the VHE spectrum is significantly softer than in the data. To illustrate this, we show in Fig.~\ref{fig:modelling_test_delta30} a modelling of the day MJD~60291 (simultaneous MAGIC/\textit{XMM-Newton} observation) using $\delta=30$ instead of $\delta=60$. We used here the same parameter values given in Tab.~\ref{tab:ssc_parameters_fixed} and Tab.~\ref{tab:ssc_parameters_evolving}, and only modified the magnetic field, electron densities, $\gamma'_{\rm max}$ and $\gamma'_{\rm min}$, to correct for the change in $\delta$. As one can see, the model is significantly softer at VHE with respect to the data, and a similar result is obtained when we modify even more parameters of the model (such as the radius).\par 

The main underlying physical reason to explain a $\delta$ as high as 60 is that in a leptonic scenario we have $\delta \propto B' \, \nu_{IC}^2/\nu_{s}$, where $\nu_{IC}$ and $\nu_{s}$ are the observed inverse-Compton and synchrotron peak frequencies. However, during December 2023, the rather soft X-ray spectra combined with a VHE hardness closer to typical states (see Fig.~\ref{fig:HR_xray_vhe}) leads to a atypically large $\nu_{IC}^2/\nu_{s}$ ratio. Hence one needs a relatively higher $\delta$, higher than usual, to reproduce the large separation of the components.

\end{appendix}

\end{document}